\documentclass[5p,twocolumn,10pt]{elsartpublic}
\journal{Computer-Aided Design}

\usepackage{natbib}

\usepackage{graphicx,amssymb,amscd,lineno,subfigure}
\usepackage{algorithm}
\usepackage{amsmath}
\usepackage[noend]{algpseudocode}
\usepackage{overpic} 
\usepackage{epstopdf} % needed if you have eps figures in a pdflatex manuscript
\usepackage{mathptmx} % use Times fonts if available on your TeX system
\usepackage{color,url}

\usepackage{tikz}\usetikzlibrary{shapes,arrows}
\usepackage{listings}
\usepackage{natbib}

\definecolor{drot}{rgb}{0.7,0,0.1}
\definecolor{ggreen}{rgb}{0.1,0.6,0.1}
\newcommand{\mk}[1]{{\color{drot} M2K: #1}}
\newcommand{\mg}[1]{{\color{drot} M2Gershon: #1}}

\usepackage{bm} %%for bold Greek letters

\usepackage{todonotes}

\newdefinition{rem}{Remark}
\usepackage[utf8]{inputenc}
\usepackage{tikz}
\definecolor{ccccccc}{RGB}{204,204,204}
\definecolor{cffffff}{RGB}{255,255,255}
\definecolor{cff0000}{RGB}{255,0,0}
\definecolor{c0000ff}{RGB}{0,0,255}
\definecolor{c00ff00}{RGB}{0,255,0}

\definecolor{blau}{rgb}{0.15,0.2,0.5}
\definecolor{drot}{rgb}{0.7,0,0.1}

\usepackage[all]{xy}

\usepackage{algorithm,algpseudocode}

\usepackage{natbib}

\newcommand{\f}[1]{\mathbf{#1}}

\newcommand{\mm}{\mathbf m}

\newcommand{\Bezier}{B\'{e}zier}

\newcommand{\Bspline}{B-spline}

\begin{document}

%\tableofcontents
%\clearpage

\begin{frontmatter}

\title{On design, analysis, and hybrid manufacturing of microstructured blade-like geometries}

\author[EPFL]{Pablo Antolin}
\ead{pablo.antolin@epfl.ch}

\author[BCAM,Ikerbasque]{Michael Barto\v{n}\corref{cor1}}
\ead{mbarton@bcamath.org}

\author[INRIA]{Georges-Pierre Bonneau}
\ead{Georges-Pierre.Bonneau@inria.fr}

\author[EPFL]{Annalisa Buffa}
\ead{annalisa.buffa@epfl.ch}

\author[UPV]{Amaia Calleja-Ochoa}
\ead{amaia.calleja@ehu.eus}

\author[Technion]{Gershon Elber}
\ead{gershon@cs.technion.ac.il}

\author[TUW]{Stefanie Elgeti}
\ead{stefanie.elgeti@tuwien.ac.at}

\author[UPV]{Gaizka G\'{o}mez Escudero}
\ead{gaizka.gomez@ehu.eus}

\author[Trimek]{Alicia Gonzalez}
\ead{agonzalez@innovalia.org}

\author[UPV]{Haizea Gonz\'{a}lez Barrio}
\ead{haizea.gonzalez@ehu.eus}

\author[INRIA]{Stefanie Hahmann}
\ead{stefanie.Hahmann@inria.fr}

\author[UTBM,EPFL]{Thibaut Hirschler}
\ead{thibaut.hirschler@utbm.fr}

\author[Technion,Hanyang]{Q Youn Hong}
\ead{qhong83@hanyang.ac.kr}

\author[TUW]{Konstantin Key}
\ead{konstantin.key@tuwien.ac.at}

\author[SNU]{Myung-Soo Kim}
\ead{mskim@snu.ac.kr}

\author[TUW]{Michael Kofler}
\ead{michael.kofler@tuwien.ac.at}

\author[UPV]{Norberto Lopez de Lacalle}
\ead{norberto.lzlacalle@ehu.eus}

\author[Trimek]{Silvia de la Maza}
\ead{smaza@trimek.com}

\author[BCAM]{Kanika Rajain}
\ead{kanika@bcamath.org}

\author[TUW]{Jacques Zwar}
\ead{jacques.zwar@tuwien.ac.at}

\cortext[cor1]{Corresponding author}

\address[Technion]{Computer Science Department, Technion -- Israel Institute of Technology, Haifa, Israel} 
\address[SNU]{Department of Computer Science and Engineering, Seoul National University, Seoul, Republic of Korea} 
\address[INRIA]{Univ.\ Grenoble Alpes, Inria, CNRS, Grenoble INP, LJK, Grenoble, France}  
\address[TUW]{Institute of Lightweight Design and Structural Biomechanics, Faculty of Mechanical Engineering and Management Sciences, TU Wien, Vienna, Austria}
\address[EPFL]{Institute of Mathematics, \'Ecole Polytechnique F\'ed\'erale de Lausanne, Switzerland}
\address[Trimek]{Trimek, Innovalia Metrology, Camino de la Yesera 2, 01139 Altube-Zuia,\'{A}lava, Spain,}
\address[UPV]{High Performance Manufacturing Group, Department of Mechanical Engineering,\\ 
the University of the Basque Country, Plaza Ingeniero Torres Quevedo 1, 48013, Bilbao, Basque Country, Spain}
\address[BCAM]{BCAM -- Basque Center for Applied Mathematics,  Alameda de Mazarredo 14, 48009 Bilbao, Basque Country, Spain}    
\address[Ikerbasque]{Ikerbasque -- Basque Foundation for Sciences, Maria Diaz de Haro 3, 48013 Bilbao, Basque Country, Spain}  
\address[UTBM]{Universit\'e de Technologie de Belfort-Montb\'eliard, France} 
\address[Hanyang]{Department of Computer Science and Engineering, Hanyang University, Ansan, Republic of Korea}

\begin{abstract}

%We aim at new generation of CAD objects that are not solid, but contain heterogeneous free-form internal microstructures.

With the evolution of new manufacturing technologies such as multi-material 3D printing, one can think of new type of objects that consist of considerably less, yet heterogeneous, material, consequently being porous, lighter and cheaper, while having the very same functionality as the original object when manufactured from one single solid material. We aim at questioning five decades of traditional paradigms in geometric CAD and focus at new generation of CAD objects that are not solid, but contain heterogeneous free-form internal microstructures. We propose a unified manufacturing pipeline that involves all stages, namely design, optimization, manufacturing, and inspection of microstructured free-form geometries. We demonstrate our pipeline on an industrial test case of a blisk blade that sustains the desired pressure limits, yet requires significantly less material when compared to the solid counterpart.  
\end{abstract}

\begin{keyword}
heterogeneous microstructure \sep microstructured free-forms \sep shape optimization \sep isogeometric analysis \sep hybrid manufacturing \sep  laser bed power fusion \sep 5-axis CNC machining \sep finishing operations
\end{keyword}

\end{frontmatter}
% \linenumbers

%%%%%%%%%%%%%%%%%%%%%%%%%%%%%%%%%%%%%%%%%%%%%%%%%%%%%%%%%%%%%
\section{Introduction}\label{sec:Intro}
%%%%%%%%%%%%%%%%%%%%%%%%%%%%%%%%%%%%%%%%%%%%%%%%%%%%%%%%%%%%%

\begin{figure}[!tbh]
\vrule width0pt \hfill
  \begin{minipage}[c]{0.48\columnwidth}
    \begin{overpic}[width=0.98\textwidth]{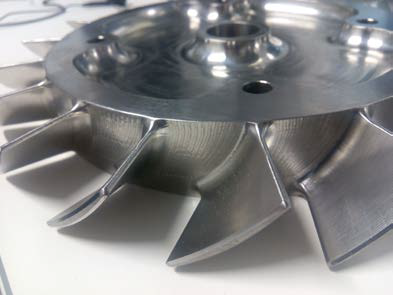}
    \end{overpic}
  \end{minipage}\hfill
  \begin{minipage}[c]{0.4\columnwidth}
  \vspace{1pt}
 \caption{A blade integrated disk, aka blisk, is a typical aeronautical workpiece that is 5-axis CNC machined from a single material block. The whole blisk, including curved, free-form blades is usually solid which makes it a heavy component of an aero-engine.}\label{Fig:Blisk}
  \end{minipage}
  \hfill \vrule width0pt
\vspace{-5ex}
\end{figure}

Curved aka free-form objects appear in various industries, aeronautical sector being a prime example where light-weight, yet strong and thermally resistant, objects are of major importance. Note that one kilogram flying implies 45K litres of fuel in an airplane lifespan, or stated differently, saving only 1$\%$ of the aircraft weight would reduce fuel consumption by 2.17 million tons per year~\cite{AeroFuelData}. Therefore one aims to reduce the weight of an aicraft wherever possible, from light seats, over plastic compartments, up to, most importantly, the metal, heat-resistant parts, parts that are the heaviest ones.

%\begin{figure}[!tbh]
%\hfill
%  \begin{overpic}[angle=-90,width=0.48\columnwidth]{fig/Figure2a}
%  \put(-2,2){(a)}
%  \put(55,20){ $\Phi_1$}
%  \end{overpic} 
%\hfill
%  \begin{overpic}[angle=-90,width=0.48\columnwidth]{fig/Figure2b}
%  \put(-2,2){(b)}
%  %\put(55,20){ $\Phi_1$}
%  \end{overpic} 
%\hfill \vrule width0pt\\
%\vspace{-15pt}
%	\caption{Microstrucured blade-like geometry. (a) A solid blade mounted on a 3D block. The boundary of the blade is formed by a free-form surface $\Phi_1$. (b) The interior of the blade is filled by a heterogenous microstructure that reduces the amount of material. }\label{fig:TestCaseGeo}
%\end{figure}

\begin{figure*}[!tbh]
\hfill
  \begin{overpic}[angle=-90,width=0.33\textwidth]{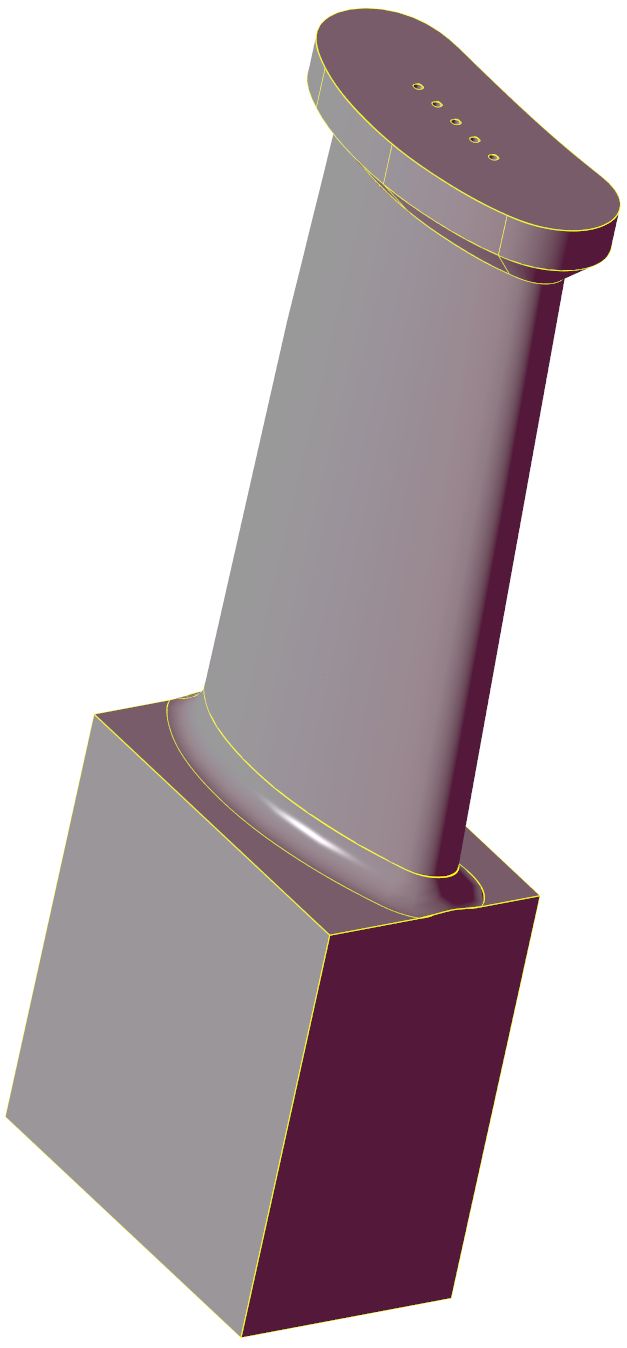}
  \put(-2,2){(a)}
  \put(55,20){ $\Phi$}
  \end{overpic} 
\hfill
  \begin{overpic}[angle=-90,width=0.33\textwidth]{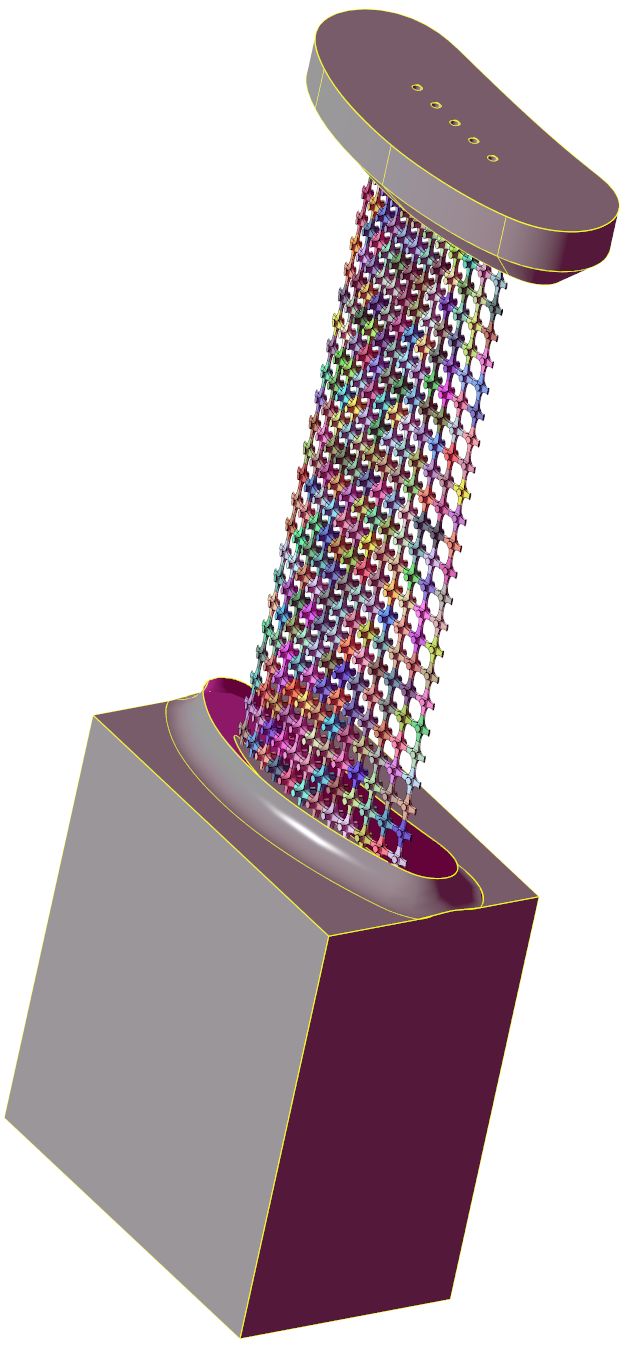}
   \put(60,50){\fcolorbox{gray}{white}{\includegraphics[angle=-90, width=0.1\textwidth]{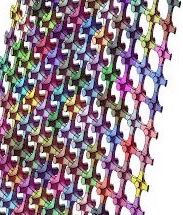}}}
  \put(-2,2){(b)}
  \end{overpic}
  \hfill
  \begin{overpic}[angle=-90,width=0.33\textwidth]{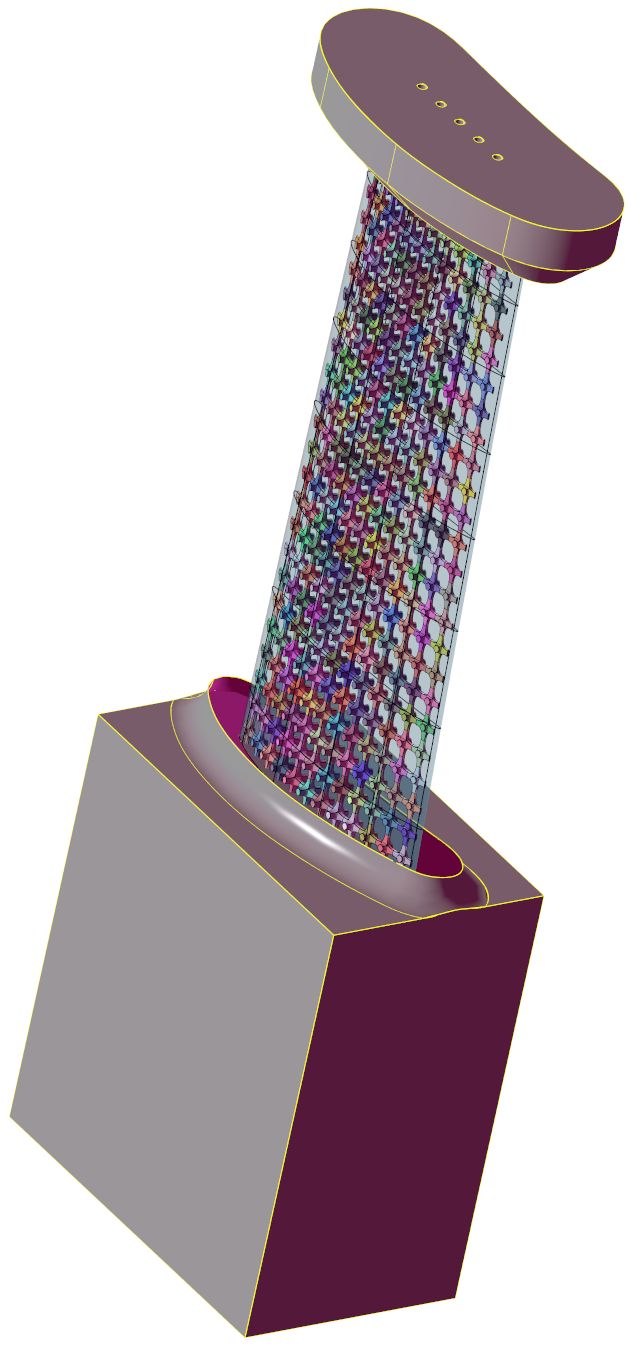}
  \put(50,32){\fcolorbox{gray}{white}{\includegraphics[angle=-90, width=0.06\textwidth]{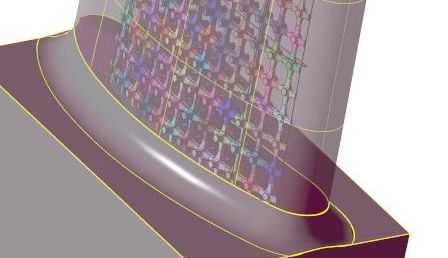}}}
  %\put(50,32){\fcolorbox{gray}{white}{\includegraphics[angle=-90, width=0.06\textwidth]{fig/Figure2e}}}
  \put(-2,2){(c)}
  \put(75,23){ $\Phi^{off}$}
  \put(60,2){ $\Phi$}
  \end{overpic} 
\hfill \vrule width0pt\\
\vspace{-32pt}
	\caption{Microstrucured blade-like geometry. (a) A solid blade mounted on a 3D block. The boundary of the blade is formed by a free-form surface $\Phi$. (b) The interior of the blade is filled by a heterogeneous microstructure that reduces the amount of material. The zoomed-in microstructure is shown framed; each tile in the microstructure is randomly colored. (c) The microstructure is surrounded by a shell $\Phi^{off}$ (transparent) and the space between the shell and $\Phi$ is filled by solid material; see transparent $\Phi$ in the framed box to feel the thickness of the material surrounding the microstructure. The five holes on the very top of the blade (right side) are designed to the removal of the support material.}
    \label{fig:TestCaseGeo}
\end{figure*}

An example of one such a curved aeronautical component, a blade integrated disk, is shown in Fig.~\ref{Fig:Blisk}, where the shape of the blades is designed towards the optimal conversion of the rotational motion (motion of the blisk) into an air flow translation in the direction of the blisk's axis. The shape of the blades, typically doubly-curved, has its functionality and is usually determined as a minimizer of a certain governing PDE that describes the object's performance and one cannot compromise it in the manufacturing stage.

The nowadays trends in modern industry go towards new technologies that allow using less material without compromising the quality of the target workpiece. With the boom of 3D printing, one can think of objects that at first glance look the same, yet inside consist of complex \emph{microstructures}, with internal cavities that contain no material, consequently being lighter and cheaper than the fully solid counterparts. Our work contributes to this trend of technological development.

To design, optimize, manufacture, and inspect such a porous object, however, introduces a decent challenge on the design-optimization-manufacturing pipeline and on the CAD modeling tools. While it is a relatively easy task for an experienced modeler to design a free-form shape, e.g.\ using CAD tools as B-spline and/or NURBS, optimizing the shape to meet a certain functionality typically goes beyond the state of the art CAD tools. The situation is even more complex when one considers objects with \emph{internal holes} (aka porosities), as these void spaces need to be properly represented to allow for a subsequent optimization. Optimizing the internal microstructure such that the whole object preserves its functionality, e.g.~the resistance towards the centrifugal force in the case of the blisk blade testcase, goes also beyond the state of the art CAD tools.  

Manufacturing of porous objects is yet another challenge that involves the accessibility issue, in the case of 3D printing, the optimal layout of the support material, its subsequent removal, and finally high-quality surface finish, that for the blade geometry is a few tens of micrometers. 

To the best of our knowledge, the proposed work is a first-of-its-kind step in this direction that introduces the whole \emph{design-optimization-manufacturing-inspection cycle} that uses microstructural design using volumetric free-from volumetric representations (V-reps), that are compatible with the state of the art isogeometric optimization framework, followed by laser powder bed fusion (LPBF) combined with 5-axis CNC machining to achieve high-quality surface finish, and finally inspected by a tomograph to check the manufacturing quality of the internal microstructure. Our pipeline is demonstrated on a blade-like aeronautical geometry, see Fig.~\ref{fig:TestCaseGeo}, that offers a sufficiently challenging free-form geometry, with clear benefits of lightweight design.

%%%%%%%%%%%%%%%%%%%%%%%%%%%%%%%%%%%%%%%%%%%%%%%%%%%%%%%%%%%%%
\section{Related Work}\label{sec:RelWork}
%%%%%%%%%%%%%%%%%%%%%%%%%%%%%%%%%%%%%%%%%%%%%%%%%%%%%%%%%%%%%

%\mb{ TODO - Gershon, Annalisa, Norberto, Muyng-Soo. (if needed, the references from the proposal are in ADAM2.bib}

\textit{Design.} 
Design of porous and lattice-like structures goes along with the blossoming of additive
manufacturing technology. These structures exist and are designed using specific type of micro-elements with typically minimal levels of control over physical properties of the designed objects, see \cite{Elber16, Elber23, Feng18, Tang15, Nguyen16}.  An example of an optimized lattice toward temperature regulation by controlling tiles' thicknesses can be found in~\cite{Zwar2022}.
As another
example, the micro-elements can be minimal surfaces with fixed topology \cite{Feng18} and only some heuristics are
used to change parameters to control the shape of the micro-element.   Minimal surfaces, and triply periodic minimal surfaces (TPMS) specifically, are typically approximated using implicit transcendental functions and are incompatible with common CAD software presentations based on B-spline surfaces.

The common solution in CAD systems nowadays is to clip a 3D axis-parallel grid of congruent implicit/parametric tiles to the desired shape, like a wing or a turbine blade.  The result can be compatible with modern CAD tools (if parametric) and consists of (trimmed) B-spline surfaces.  Yet, tiles on the boundary of the desired shape are clipped arbitrarily, clipping that can have mechanical and/or physical effects on the final model.

In~\cite{Massarwi2016} a volumetric representation (V-rep) was presented that exploits B-spline functions. 
In~\cite{Elber16},  a lattice construction scheme was presented based on V-reps that employs functional composition to embed B-spline tiles in 3-space.  The use of functional composition in V-reps allows the representation of conformal lattice structures to the desired shape while no tile is being clipped, the tiles can consist of functionally graded materials (FGM) or be heterogeneous, and finally, the results can be fully compatible with common CAD software presentations, synthesizing B-spline surfaces, as well as full compatibility with iso-geometric analysis~\cite{igaBook09}, synthesizing B-spline trivariates.

%The European market has some limited use of microstructures too, typically in a grid-like (or lattice-like) arrangements, see for example
%\cite{LatticeURL}. 

%However, their manufacturing is a complex procedure that is accomplished
%using a specific sequence of steps that are associated with one particular type of microstructure, leading to one
%particular manufacturing algorithm. In contrast, ADAM2 proposes a unified methodology that will further
%strengthen the use of microstructures in Europe. To the best of our knowledge, the optimal layout of the internal
%parameterized micro-elements in freeform microstructure, the ability to design and to analyze them, as well as
%its manufacturing using hybrid machining (3D printing and 5-axis CNC machining) is radically new. Consequently,
%ADAM2 offers a geometric modeling framework for the 21st century that will allow for design, analysis,
%optimization, and manufacturing of highly complex porous and microstructured heterogeneous geometries.

%\SGPc{What follows is a general list of related works on auxetics. Once the other related works have been written concerning the different aspects of this paper, we will have to reduce the reference to auxetics, as it is not the central micro-structure of this publication.} 

\textit{Auxetic structures.} 
Auxetic structures are metamaterials with a negative Poisson's ratio, meaning they expand perpendicular to the direction of the applied force when stretched.
This property is unusual and may seem counter-intuitive at first glance.
Natural auxetic materials are rare, found in crystalline structures, human tendons, and the skin of snakes. In 1987, Lakes \cite{Lakes1987} led the way in auxetic material research by creating the first practical auxetic foam structure.
A significant portion of research is focusing on the design of auxetic mechanical metamaterials, which derive their extraordinary properties from small-scale, rationally designed geometric structures rather than from the material of which they are made.
These structures come in various geometric patterns, including re-entrant designs, rotating rigid forms, chiral configurations, and perforated sheets, which exhibit auxetic behavior when subjected to external loads. 
Reviews about auxetic structures in 2D and 3D can be found in \cite{Ren2018,Kolken2017,Liu2010,Greaves2011,Novak2016,Saxena2016}.
Elipe et al.\ \cite{Elipe2012} present a comparative study of most
known 2D and 3D auxetic geometries using CAD modeling, tetrahedral
meshing and FEM simulation.\\
Recent research on computational design of microstructures encompasses generating databases for discovering new 3D auxetic structures \cite{Zhou2017}, optimizing for customized mechanical properties \cite{Schumacher2018}, investigating tileable and printable designs with manufacturing constraints \cite{Panetta15}, creating gradable microstructures \cite{Martinez2016_proceduralVoronoi}, and developing disordered random auxetic structures \cite{Bonneau2021_disorderedAuxetics,Martinez2021_porousRandom}.\\
Auxetic materials find diverse applications due to their excellent energy absorption, identation resistance, synclasticity, and vibration damping properties
\cite{Saxena2016,Ren2018}. Industrial applications include 
protective gear and sportswear \cite{Duncan2018_sport},
textiles \cite{Wang2014_textile},
%sealing and gaskets,
filters and membranes \cite{Attard2018_filtering},
aerospace engineering \cite{Alderson2007_auxeticAerospace,Lira2011_fanBlade},
bio-medical devices \cite{Lvov2022_BioMed}.

\textit{Analysis on microstructures.} 
The analysis of microstructured geometries presents unique challenges that are not typically encountered in solid models. One major difficulty is the significantly higher number of discretization points required to fully capture the details of the microstructure. To mitigate the associated computational costs, various approaches have been proposed in the literature. 
One widely used strategy is homogenization, particularly when the microstructure exhibits distinct multi-scale characteristics. Originally introduced by Bensoussan et al. (1979) \cite{Bensoussan1979}, homogenization has since been applied to topology optimization, as demonstrated by Bendsoe and Kikuchi (1988) \cite{BENDSOE1988197}. Since then, homogenization has been applied in different applications and contexts, for example reviewed in \cite{encyclopedia2020072}. In particular, it was shown that the scale factor does not have to be as pronounced as originally expected \cite{coelho2016scale,CARBONARO2024108467}.  
If homogenization is not applicable, beam models can offer a useful alternative to reduce computational costs, particularly when the microstructure consists of slender structures with minimal local variation. Examples of this approach can be found in \cite{weeger2021numerical, liu2021equivalent}. 
However, when the scale factor is too small or the microstructure cannot be adequately represented by a beam model, fully resolved geometrical models may become necessary. In such cases, efficient assembly of stiffness matrices is crucial to maintaining computational feasibility \cite{hirschler2022fast}. 

\textit{Optimization of microstructures.}
In the field of microstructures, there exists a wide range of parameters that can be optimized to improve performance and functionality. These include geometric parameters at both the micro and macro scales, such as shape, orientation, and arrangement of structural features. Additionally, material parameters—such as stiffness, density, or thermal conductivity—play a crucial role in determining the overall behavior of the structure. Furthermore, topological modifications of the microstructure, such as introducing or rearranging voids, connections, or layers, can lead to significant enhancements in mechanical or functional properties. Together, these variables offer a rich design space for engineering high-performance materials and structures. There are several approaches to controlling this large design space, e.g., multi-scale optimization {\cite{xia2017recent}}, density-based homogenization methods \cite{lynch2018design}, or parameter-based methods \cite{geoffroy2017optimization}.

\textit{CNC machining of blade-like geometries.} 
%\cite{Zaragoza-2024-Collision,Rajain-2025-G1conicalPhysical,Martinez-2024-TurbomachineryBlade,Calleja-2019-Blisk}
Machining of blades, particularly those used in high-performance applications like turbines, requires high precision to achieve optimal aerodynamic behavior. Traditional subtractive manufacturing techniques are often employed to refine blade geometries, ensuring tight tolerances and smooth surfaces that enhance performance and longevity~\cite{Rajain-2025-G1conicalPhysical,Martinez-2024-TurbomachineryBlade,Calleja-2019-Blisk}. Moreover, the blade-like geometries contain cavities which poses another challenge in terms of collision avoidance~\cite{Zaragoza-2024-Collision}.  

With the advent of additive manufacturing, specifically laser powder bed fusion (LPBF), the production of complex blade geometries has become more feasible. LPBF enables the fabrication of intricate designs that are challenging for conventional methods. However, the as-built surfaces from LPBF processes typically exhibit roughness and residual stresses, necessitating post-processing treatments. Studies have shown that techniques like centrifugal disk finishing can effectively smooth the surfaces of LPBF-manufactured components, such as Inconel 718 blades, without significantly altering the bulk microstructure. This process not only improves surface finish but also imparts beneficial residual stress profiles, enhancing the mechanical performance of the blades \cite{Peterson-2024-UPV1ref}.
Moreover, the LPBF process can induce unique microstructural features due to rapid solidification rates. For instance, LPBF can produce bulk nanostructured alloys with interconnected phase networks, which may contribute to improved mechanical properties~\cite{Jung-2020-UPV2ref}.
%
%In summary, while LPBF offers significant advantages in manufacturing complex blade geometries, careful consideration of post-processing techniques is essential to achieve the desired surface quality and mechanical properties.

\begin{figure*}[!tbh]
\centering
\hfill
  \begin{overpic}[width=0.46\textwidth]{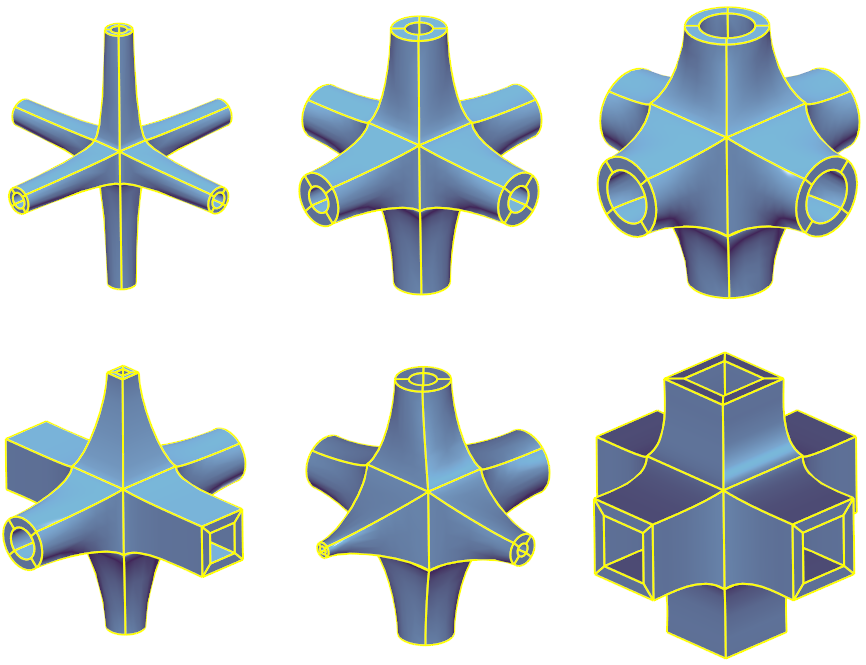}%,grid,tics=20
   %\put(0,40){\fcolorbox{gray}{white}{\small{$p=??$}}}
  		\put(0,0){(a)}
  \end{overpic}\hfill
  \begin{overpic}[width=0.46\textwidth]{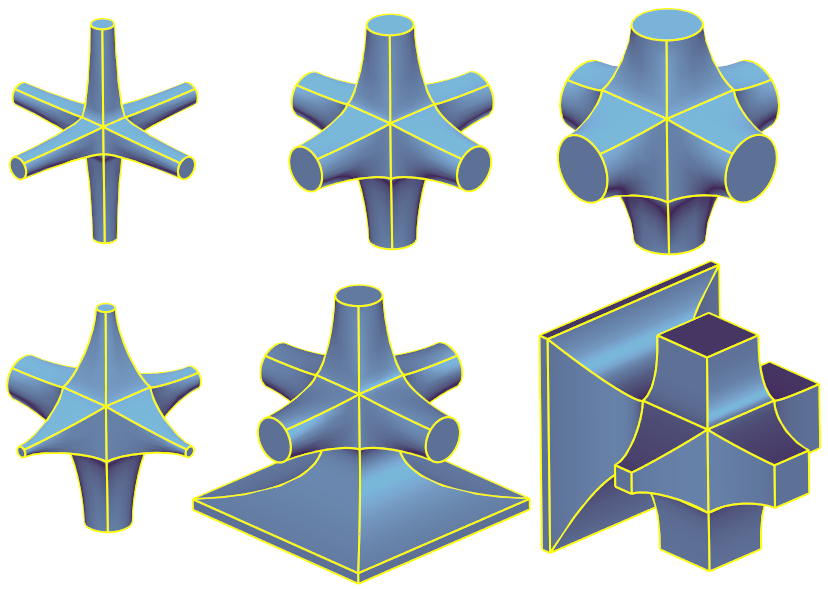}%,grid,tics=20
  		\put(0,0){(b)}
  \end{overpic}\hfill
  \hfill \vrule width0pt\\
 \vspace{2pt}
	\caption{Gallery of cross tiles controlled via a family of parameters. (a) A set of six microtiles containing interior holes are shown. One can control the thickness of the pipes as well as their roundness. (b) The analogy of tiles with no internal holes, including two special tiles that support an attachment feature to the boundary shell (bottom right).}\label{fig:GalleryOfTiles}
\end{figure*}

\begin{figure}[!tbh]
\begin{center}
\hfill
\begin{overpic}[width=0.45\columnwidth]{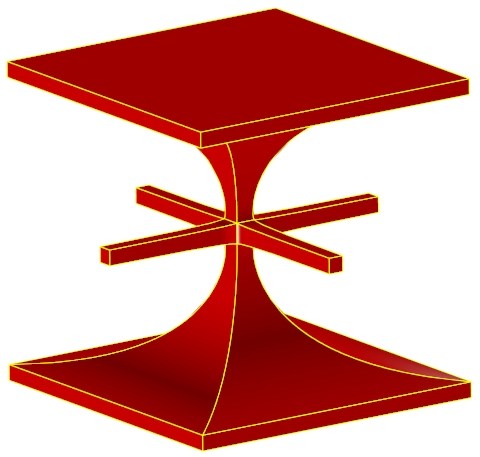}
\put(-5,0){(a)}
\end{overpic} 
\hfill
\begin{overpic}[width=0.45\columnwidth]{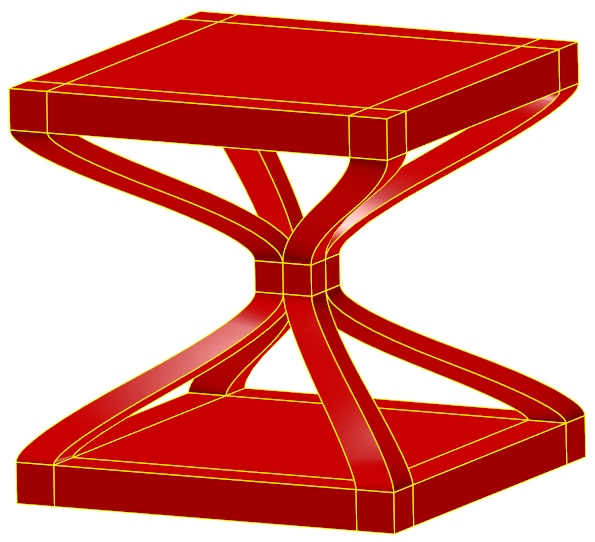}
\put(-5,0){(b)}
\end{overpic} 
\hfill \vrule width0pt\\
%\begin{picture}(0,0)
%    \put( -230, 152){(a)}
%    \put(  -35, 153){(b)}
%\end{picture}
\end{center}
\vspace{-16pt}
\caption{A cross-like tile with axis parallel arms is shown in (a) and a diagonal-arms tile in (b).  Both with surface skin. These tiles are samples from a continuous family of tiles, controlled via several parameters such as arm and skin thickness, roundness, etc.}\label{fig-blade-lattice-tiles}
\end{figure}

%%%%%%%%%%%%%%%%%%%%%%%%%%%%%%%%%%%%%%%%%%%%%%%%%%%%%%%%%%%%%
\section{Volumetric microstructural design}\label{sec:Design}
%%%%%%%%%%%%%%%%%%%%%%%%%%%%%%%%%%%%%%%%%%%%%%%%%%%%%%%%%%%%%

In this Section, we discuss two particular designs of internal microstructures. First, we consider design using repetitive, yet heterogeneous, rigid lattices in Section~\ref{ssec:StaticMicro}. Second -- since our target geometry are blades in aeronautics -- we consider also auxetic microstructures that are meant to adapt towards centrifugal forces that will act on the blade during its functionality, in  Section~\ref{ssec:AuxetMicro}. The test case geometry under consideration is a blade, recall Fig.~\ref{fig:TestCaseGeo}. Its interior is aimed to be filled by a microstructure, however, since the workpiece needs to be CNC machined as the very final step of the manufacturing process, sufficiently thick shell (aka skin) of solid material has to be left close to the boundary surface $\Phi$. Therefore we microstructure only the interior which is delimited by the offset surface, $\Phi^{off}$, with offset distance being the width of the skin.

%\SGPc{Shall we give details about the geometric shape of the cavity to be filled with microstructures and a figure showing this cavity here?.}

%%%%%%%%%%%%%%%%%%%%%%%%%%%%%%%%%%%%%%%%%%%%%%%%%%%%%%%%%%%%%
\subsection{Constructing the free--form microstructures}\label{ssec:StaticMicro}
%%%%%%%%%%%%%%%%%%%%%%%%%%%%%%%%%%%%%%%%%%%%%%%%%%%%%%%%%%%%%

%\begin{figure}[!tbh]
%\hfill
%  \begin{overpic}[height=0.20\columnwidth]{fig/blade_cavity.png}
%  \end{overpic} 
%\hfill \vrule width0pt\\
%\vspace{-24pt}
%	\caption{\SGPc{We need figures about exterior and interior geometry, all of same style. These figures are only place holders.} \mb{I agree, we need a nice illustration of the skin, i.e, $\Phi$ and $\Phi^{off}$, Kanika.}}\label{fig:bladeGeometry}
%\end{figure}

\begin{figure}
\begin{center}
\hfill
\begin{overpic}[width=0.95\columnwidth]{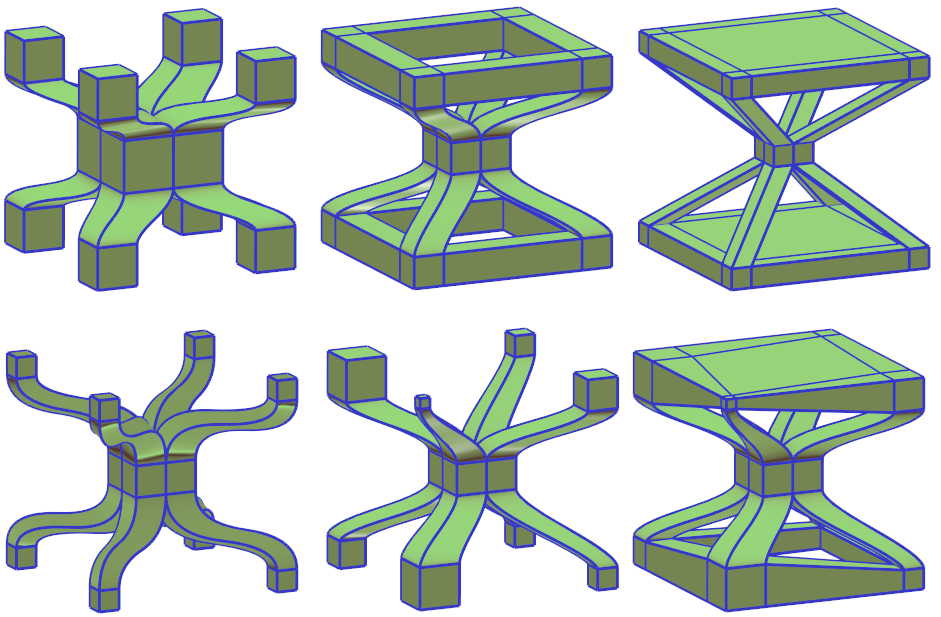}
%\put(-5,0){(a)}
\end{overpic} 
\hfill \vrule width0pt\\
%\begin{picture}(0,0)
%    \put( -230, 152){(a)}
%    \put(  -35, 153){(b)}
%\end{picture}
\end{center}
\vspace{-16pt}
\caption{Several variants of the diagonal tile controlled via a family of parameters, recall Fig.~\ref{fig-blade-lattice-tiles}(b), including tiles that attach to one (bottom right), or two opposite (top right) faces. This type of tile is later used in Fig.~\ref{fig-blade-lattice-diagonal} to attach the microstructure to the outer skin.
}\label{fig-Diagonal-Tiles}
\end{figure}

Several variations of lattices were created for the blade, to begin without outer surface skin (hence after {\em skin}), and as presented here with skin.
See Figs.~\ref{fig:GalleryOfTiles} and~\ref{fig-Diagonal-Tiles} for some examples of tiles.
The two tiles used in this effort are shown in Fig.~\ref{fig-blade-lattice-tiles}. These tiles are parametric
which means that the geometry of tiles could be adjusted within the lattice and even within individual tiles. The sizes of the tiles' arms, the skin existence and its thickness, etc., can all be adjusted
along the lattice and also within individual tiles. Again, recall Figs.~\ref{fig:GalleryOfTiles} and~\ref{fig-Diagonal-Tiles}.

The blade itself was modeled as a trivariate volumetric model and its domain was populated with the tiles, only to functionally compose these tiles with the trivariate of the blade.  More detail on this construction can
be found in~\cite{Elber23}.

The cross tile shown in Fig.~\ref{fig-blade-lattice-tiles}(a) contains \Bezier{} trivariates that are either tri-linear or of orders $(2 \times 2 \times 4)$ and $(2\times 2 \times 5)$.  The diagonal tile
shown in Fig.~\ref{fig-blade-lattice-tiles}(b) contains \Bezier{} and \Bspline{} trivariates that are either tri-linear or of orders $(3
\times 2 \times 2)$ (and control mesh size of $(4 \times 2 \times 2))$. The blade trivariate employed in this section in all presented lattices is of orders $(3 \times 2 \times 2)$ and mesh size of $(5
\times 2 \times 2)$.
%\cre{A gallery o possible tiles and their variations is shown in Fig.~\ref{fig:GalleryOfTiles}.}

\begin{figure}[!tbh]
\begin{center}
\hfill
\begin{overpic}[width=0.85\columnwidth]{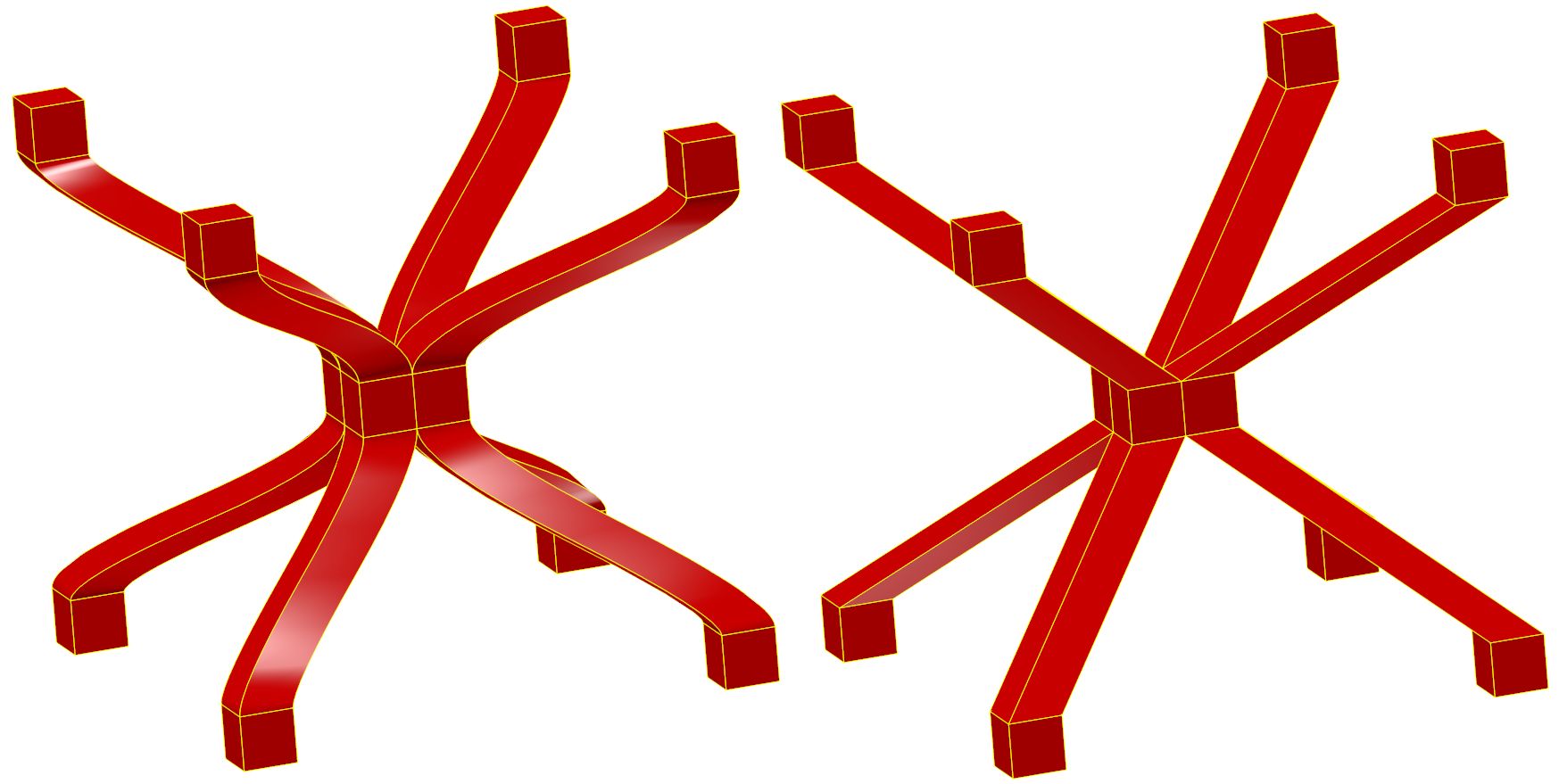}
\put(10,45){\fcolorbox{gray}{white}{MS1R}}
\put(60,45){\fcolorbox{gray}{white}{MS1D}}
\end{overpic} 
\hfill \vrule width0pt\\
\vspace{-6pt}
\caption{Two diagonal cross tiles used for physical experiments in Section~\ref{sec:Manuf}. The diagonals with rounded corners, MS1R, and without rounding, MS1D.}\label{fig:MS12DR}
\end{center}
\end{figure}

The tiles in Fig.~\ref{fig-blade-lattice-tiles} are embedded to a unit cube and are designed such that the tile either contains a whole face of the cube, or an arm intersects the face at its center and under the right angle. Such a construction guarantees that, when deformed by a smooth mapping, the neighboring tiles are smoothly joined, forming a $C^1$ continuous microstructure.

\begin{figure*}[!tbh]
\begin{center}
\begin{overpic}[width=0.49\textwidth]{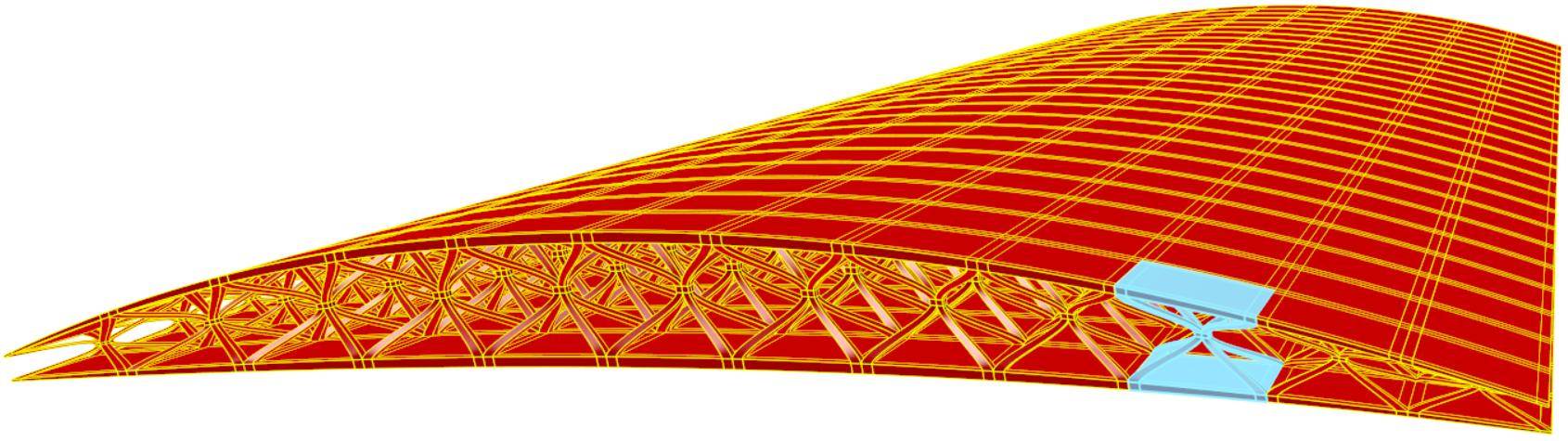}
% \put(0,13){\fcolorbox{gray}{white}{\includegraphics[width=0.1\textwidth]{fig/DiagTileSkin.jpg}}}
 \put(10,15){\fcolorbox{gray}{white}{MS1R}}
 \put(3, 23){(a)}
\end{overpic}
\begin{overpic}[width=0.49\textwidth]{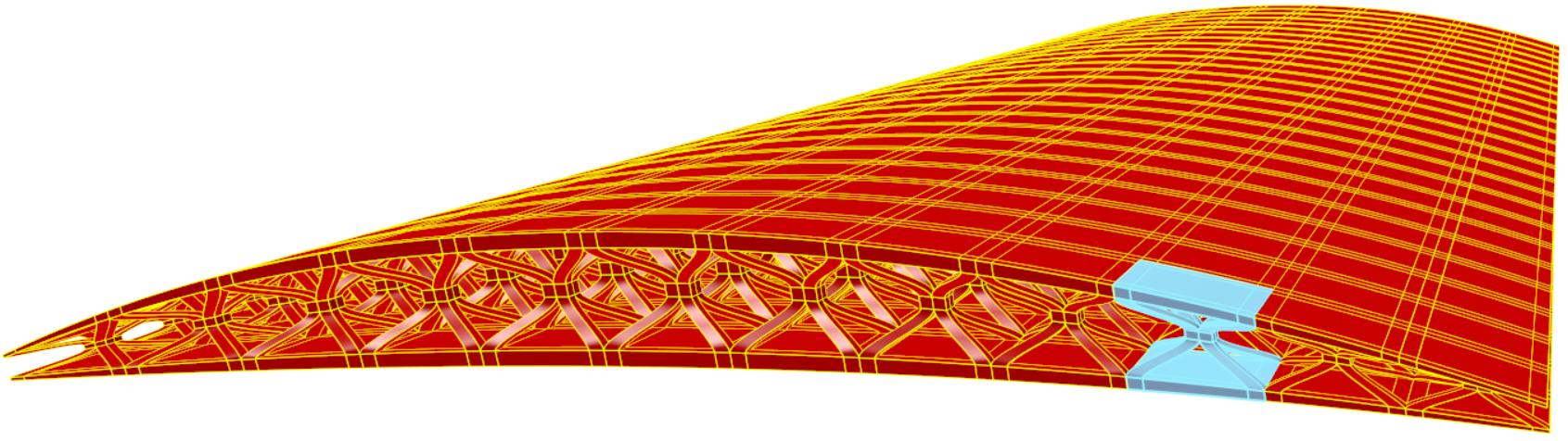}
 \put(10,15){\fcolorbox{gray}{white}{MS2R}}
 \put(35,20){$\Phi^{off}$}
 \put(3, 23){(b)}
\end{overpic}
\begin{overpic}[width=0.49\textwidth]{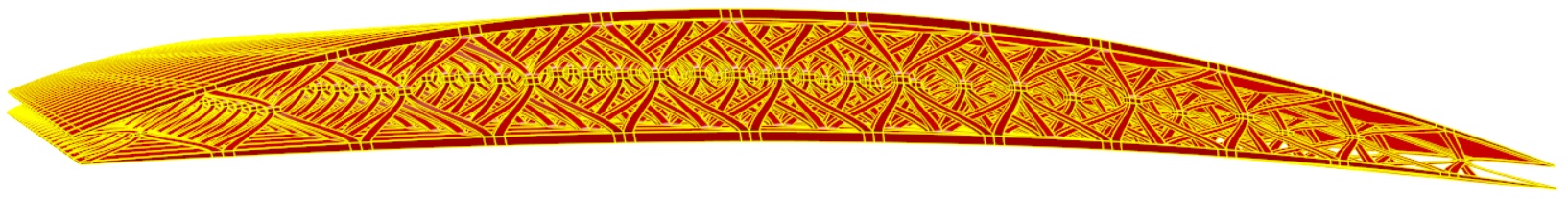}
\end{overpic}
\begin{overpic}[width=0.49\textwidth]{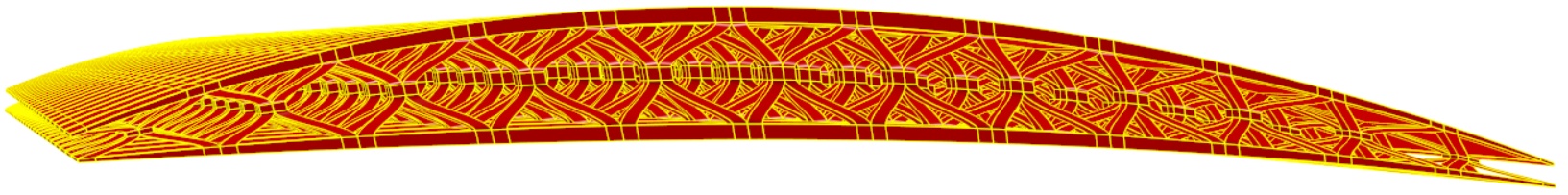}
\end{overpic}
%\begin{picture}(0,0)
%    \put( -390, -2){(a)}
%    \put( -160, -2){(b)}
%\end{picture}
\end{center}
\vspace{-16pt}
%\mbox{\vspace{-0.5in}}\\[-6pt]
\caption{Two variations of the blade lattices using the  canonical "diagonal-arms" tile (top framed); recall
 also Fig.~\ref{fig-blade-lattice-tiles}(b). The canonical tile is deformed and mapped throughout the interior of the blade, forming a heterogeneous microstructure that comforts the outer surface shell $\Phi^{off}$.}
\label{fig-blade-lattice-diagonal}
\end{figure*}

\begin{figure*}[!tbh]
\begin{center}
\begin{overpic}[width=0.32\textwidth]{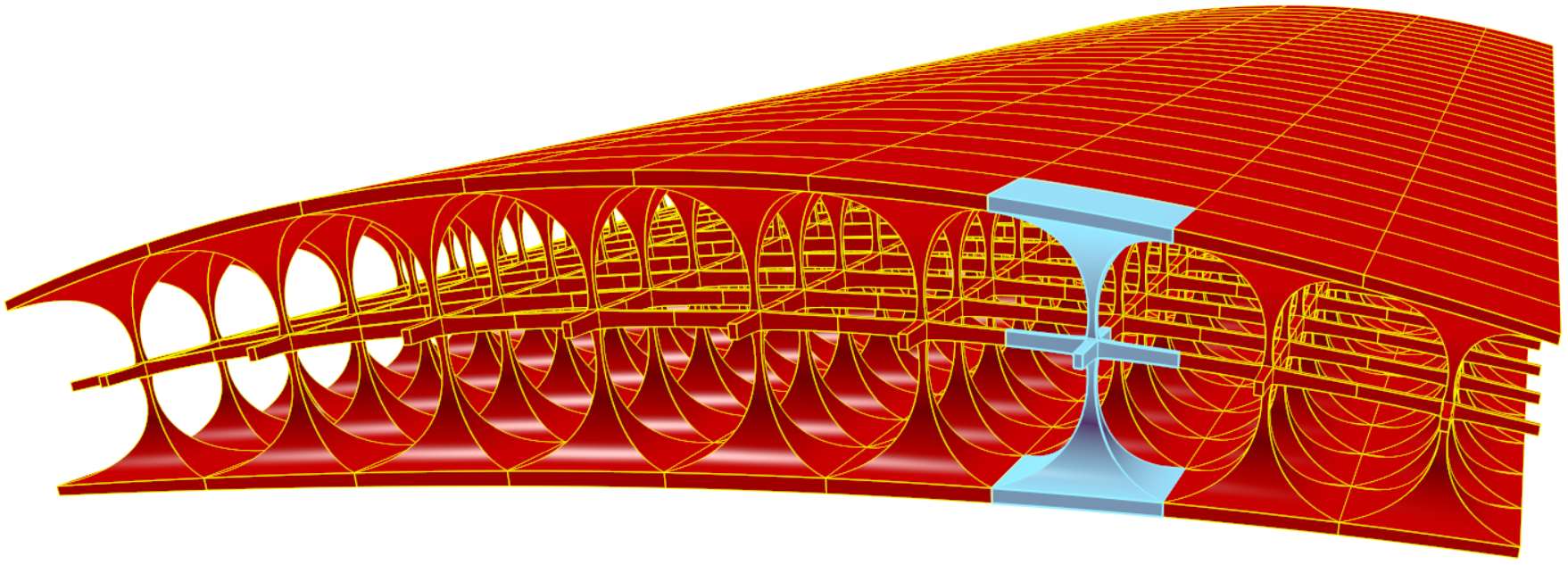}
\end{overpic}
\begin{overpic}[width=0.32\textwidth]{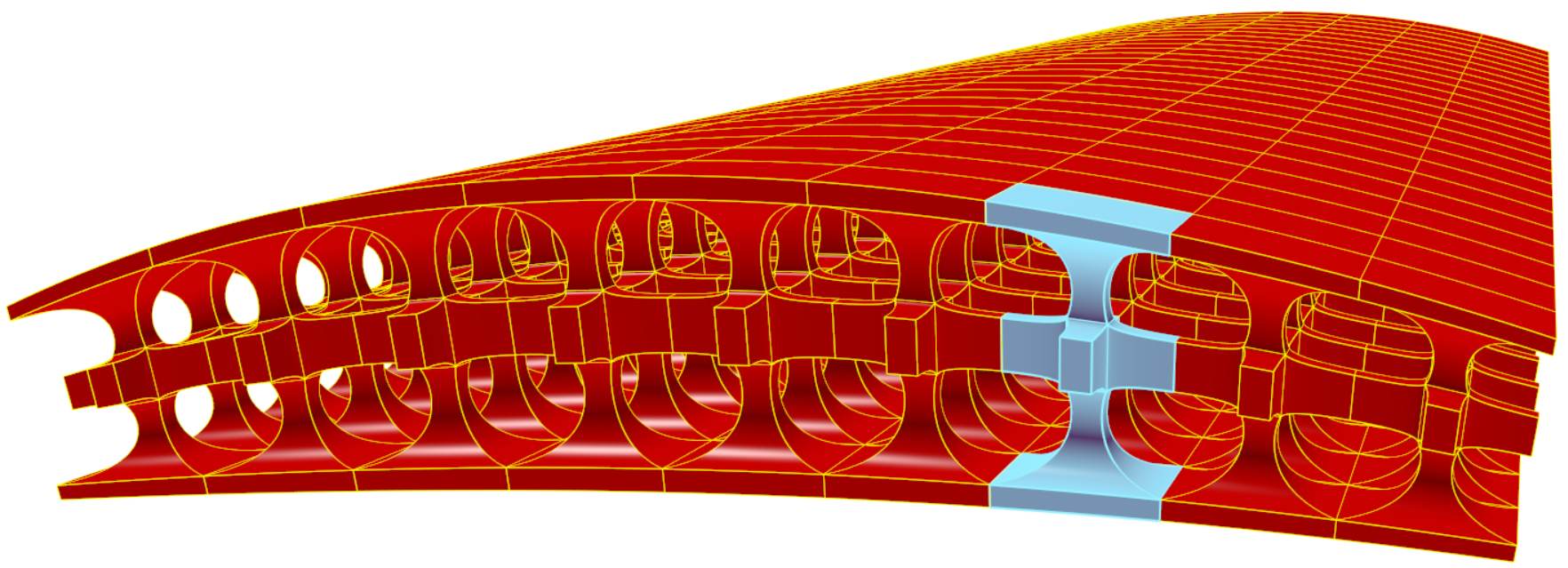}
\end{overpic}
\begin{overpic}[width=0.32\textwidth]{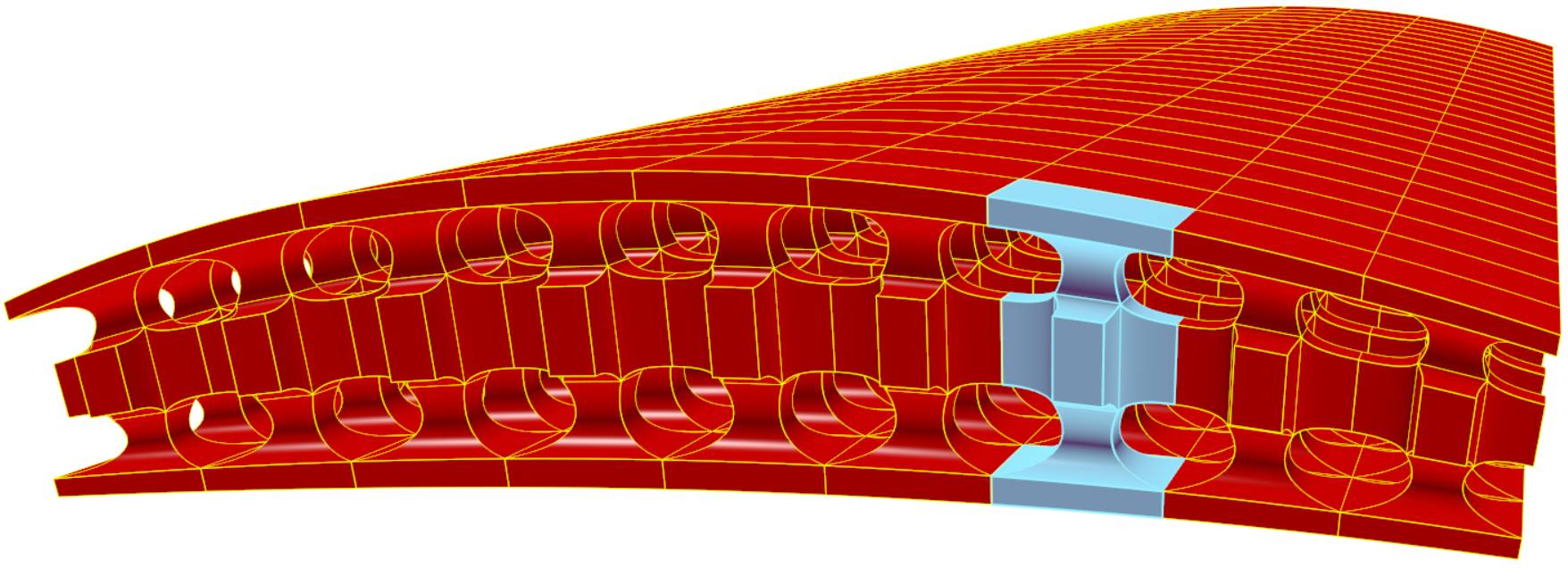}
\end{overpic}
\begin{overpic}[width=0.32\textwidth]{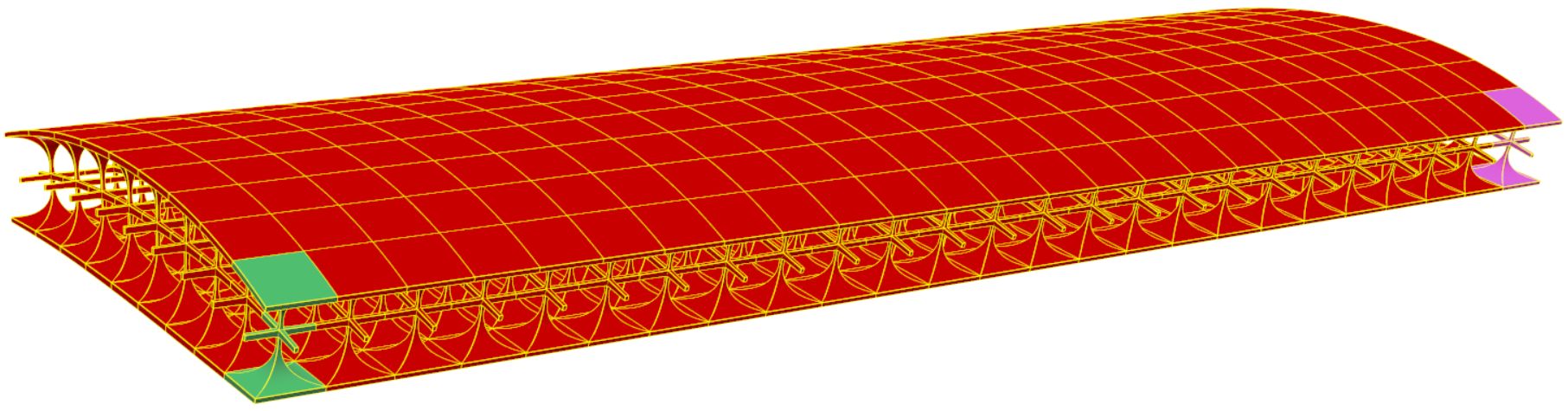}
\end{overpic}
\begin{overpic}[width=0.32\textwidth]{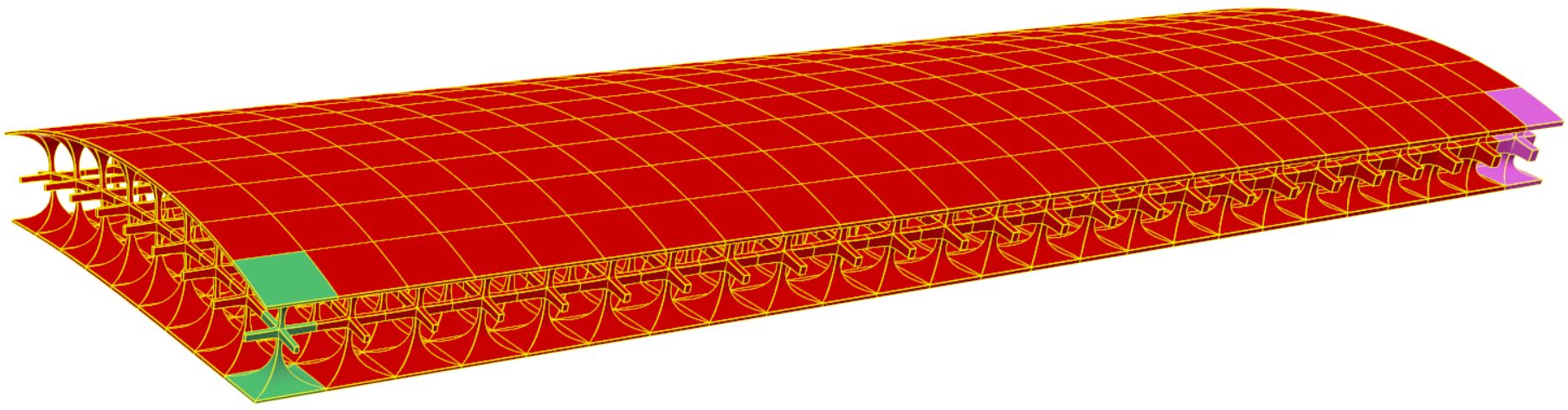}
\end{overpic}
\begin{overpic}[width=0.32\textwidth]{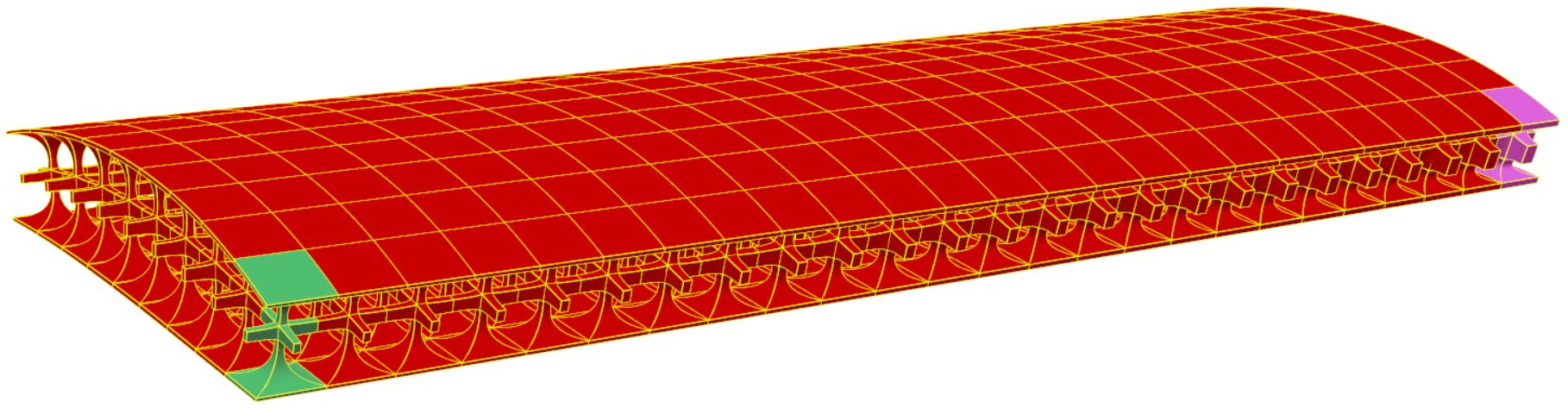}
\end{overpic}
\begin{picture}(0,0)
    \put( -380, 5){(a)}
    \put( -220, 5){(b)}
    \put(  -40, 5){(c)}
\end{picture}
\end{center}
\mbox{\vspace{-0.5in}}\\[-0.5in]
\caption{Three variations of the blade lattices with the cross tile from Fig.~\ref{fig-blade-lattice-tiles}(a).}
\label{fig-blade-lattice-cross}
\end{figure*}
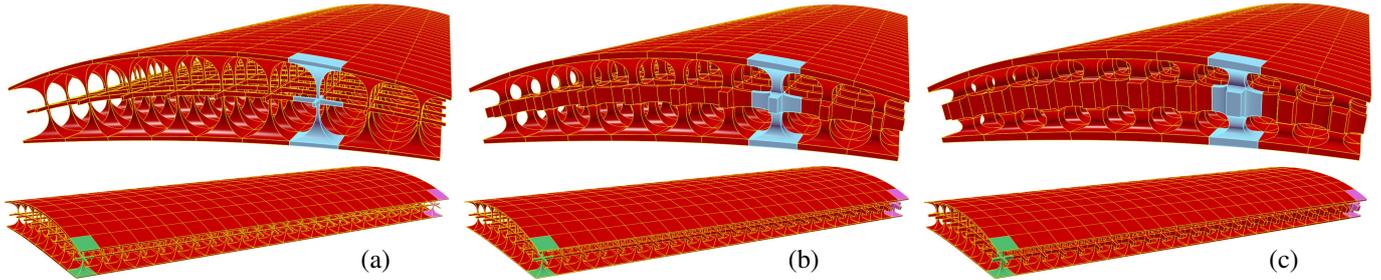

We use four types of interior diagonal tiles for the physical experiments (see later Section~\ref{sec:Manuf}); two tiles with straight diagonal edges and two with diagonal edges rounded close to the corner, see Fig~\ref{fig:MS12DR}. We also experiment with the tile thickness. That is, MS2D and MS2R are the thicker versions of MS1D and MS1R, see also  Fig.~\ref{fig-blade-lattice-diagonal} which shows the whole core lattice using the thin (MS1R) and thick (MS2R) interior tile. Here again, the tiles in these two lattices have different arm thicknesses, with one tile highlighted in cyan, at the top figures. 
%The resulting lattice contains trivariates of high orders. Hence, all trivariates in the lattice were approximated as tri-quadratics. - Gershon: already discussed in the next paragraph in even more details

\begin{figure}
\begin{center}
%\hfill
\begin{overpic}[width=0.91\columnwidth]{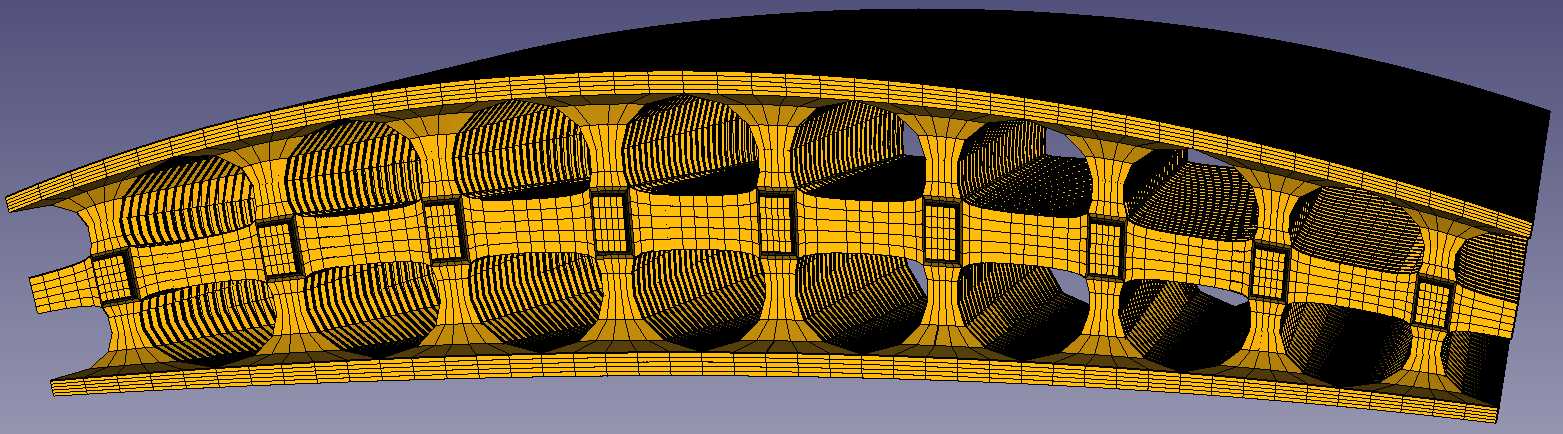}
\put(1, 24){\textcolor{white}{(a)}}
\end{overpic}
%\hfill \vrule width0pt\\
%
%\hfill
\begin{overpic}[width=0.91\columnwidth]{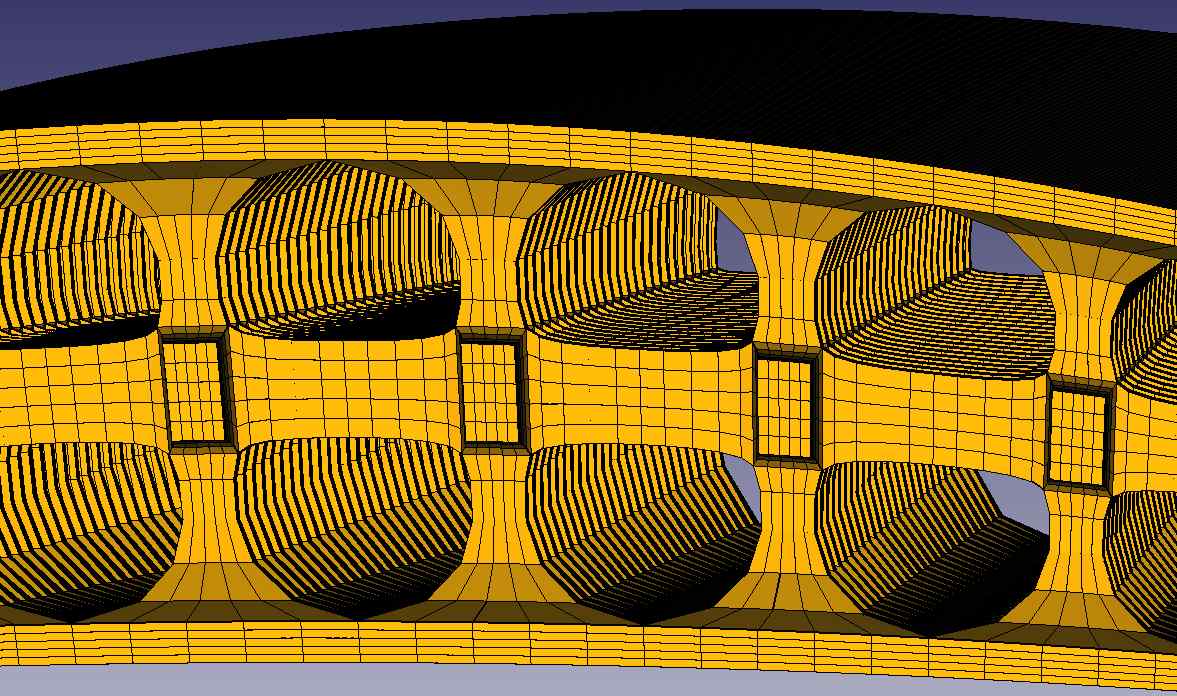}
\put(1,55){\textcolor{white}{(b)}}
\end{overpic}
%\hfill \vrule width0pt\\
%\begin{picture}(0,0)
%    \put(  -232, 192){\textcolor{white}{(a)}}
%    \put(  -232, 126){\textcolor{white}{(b)}}
%    %\put(  0, 158){\textcolor{white}{(a)}}
%    %\put(  0,  18){(b)}
%\end{picture}
\end{center}
\vspace{-15pt}
%\mbox{\vspace{-0.2in}}\\[-0.2in]
\caption{A finite element hexa (cuboid) mesh result of a lattice, ready for analysis (a). (b) A a zoom-in of (a).}\label{fig-blade-lattice-FEM}
\end{figure}

%\begin{figure}[!tbh]
%\hfill
%  \begin{overpic}[width=0.88\columnwidth]{fig/BladeWithSkin2}
%  \end{overpic} 
%\hfill \vrule width0pt\\
%\vspace{-18pt}
%	\caption{A non-homogeneous microstructure that conforms the free-form shape of the %blade.}\label{fig:NonHomoMicro}
%\end{figure}

\begin{figure}[!tbh]
\hfill
%\begin{center}  
%    \begin{overpic}[width=0.26\columnwidth]{fig/doubleArrowLessCompressedHinge.png}\end{overpic}
%    \begin{overpic}[width=0.35\columnwidth]{fig/doubleArrowRest.png}\end{overpic}
%    \begin{overpic}[width=0.38\columnwidth]{fig/doubleArrowExtendedHinge.png}\end{overpic}
%\end{center}
\begin{center}  
    \begin{overpic}[width=0.88\columnwidth]{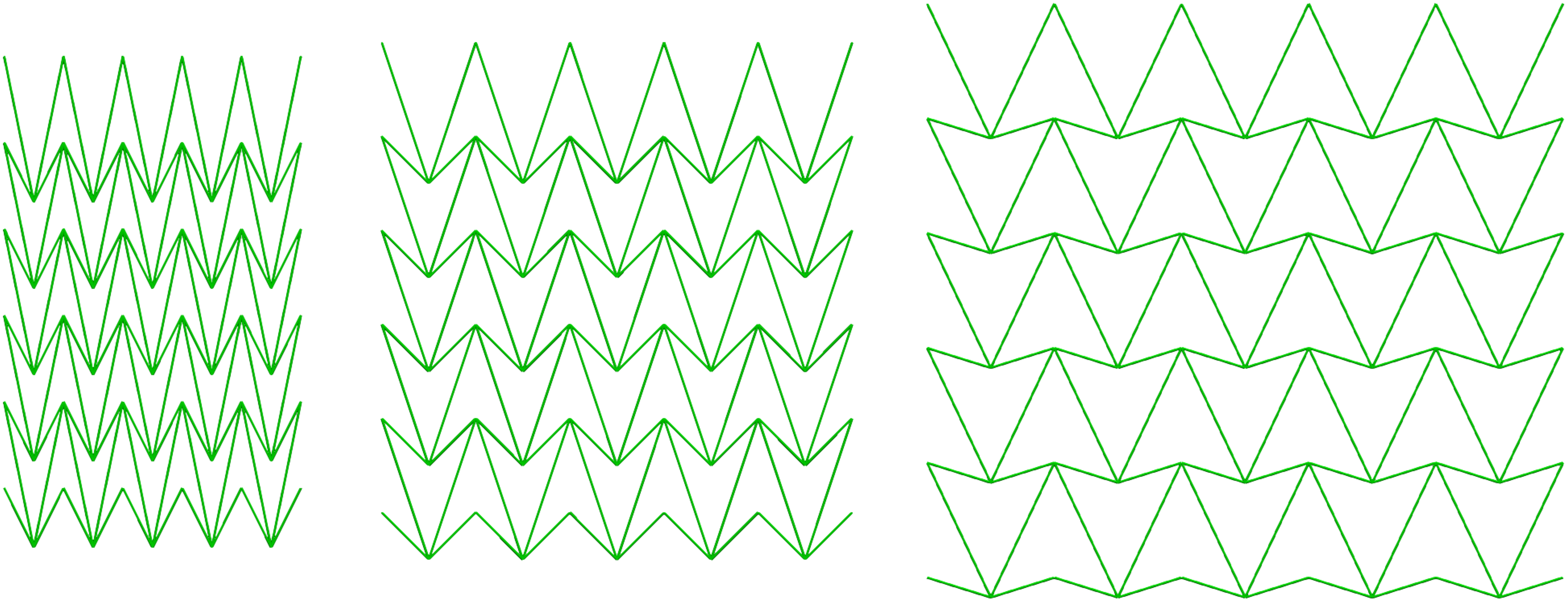}\end{overpic}
\end{center}
    %\begin{overpic}[height=0.25\columnwidth]{fig/AuxBlade.png}\end{overpic}
% \hfill \vrule width0pt\\
\vspace{-15pt}
	\caption{Auxetic behavior. A 2D double-V structure at rest (middle) shrinks vertically when compressed horizontally (left), and expanded vertically when extended horizontally (right).
     %STEF: just for information:  angles relative to the vertical of the struts are 27m 47, 73 degres}
     }
\label{fig:2DdoubleV}
\end{figure}

Fig.~\ref{fig-blade-lattice-cross} shows three lattices of the blade using cross tiles (recall Fig.~\ref{fig-blade-lattice-tiles}(a)) of different thickness from thin (a) to thick (c), with one tile highlighted in cyan, at the top.  Further, and while in (a) all tiles are of similar arm and skin thicknesses, in (b), and more so in (c), the sizes of the arms and skin continuously vary from the root of the blade (with thick tiles' geometries - see rightmost tile highlighted in magenta) to the tip of the blade (with thin tiles' geometries - see leftmost tile highlighted in green).  The resulting lattice contains trivariates of high orders due to the functional composition process, while in some cases, higher-order splines can be undesired, for example in IGA~\cite{igaBook09}.  Further, because individual tiles are typically piecewise-smooth and similarly the desired shape of the lattice is also typically locally smooth, lower-order splines can yield a good approximation. In this work, all trivariates in the lattice were approximated as tri-cubics, towards IGA.

All the geometry presented in this section is fully compatible with iso-geometric analysis (IGA, see~\cite{igaBook09}).  Further,
Fig.~\ref{fig-blade-lattice-FEM} shows an example of converting
these lattices into hex (cuboid) finite elements, with (b) a zoom-in of (a), presented in the FreeCAD CAD
software\footnote{https://www.freecad.org/}.

%%%%%%%%%%%%%%%%%%%%%%%%%%%%%%%%%%%%%%%%%%%%%%%%%%%%%%%%%%%%%
\subsection{Auxetic microstructures}\label{ssec:AuxetMicro}
\begin{figure}[!tbh]
\hfill
\begin{center}
    %\begin{overpic}[width=0.4\columnwidth]{fig/oneCellFullWithoutSupport.png}\end{overpic}
    %\begin{overpic}[width=0.4\columnwidth]{fig/oneCellWithinWallsLight.png}\end{overpic}
    \begin{overpic}[width=0.99\columnwidth]{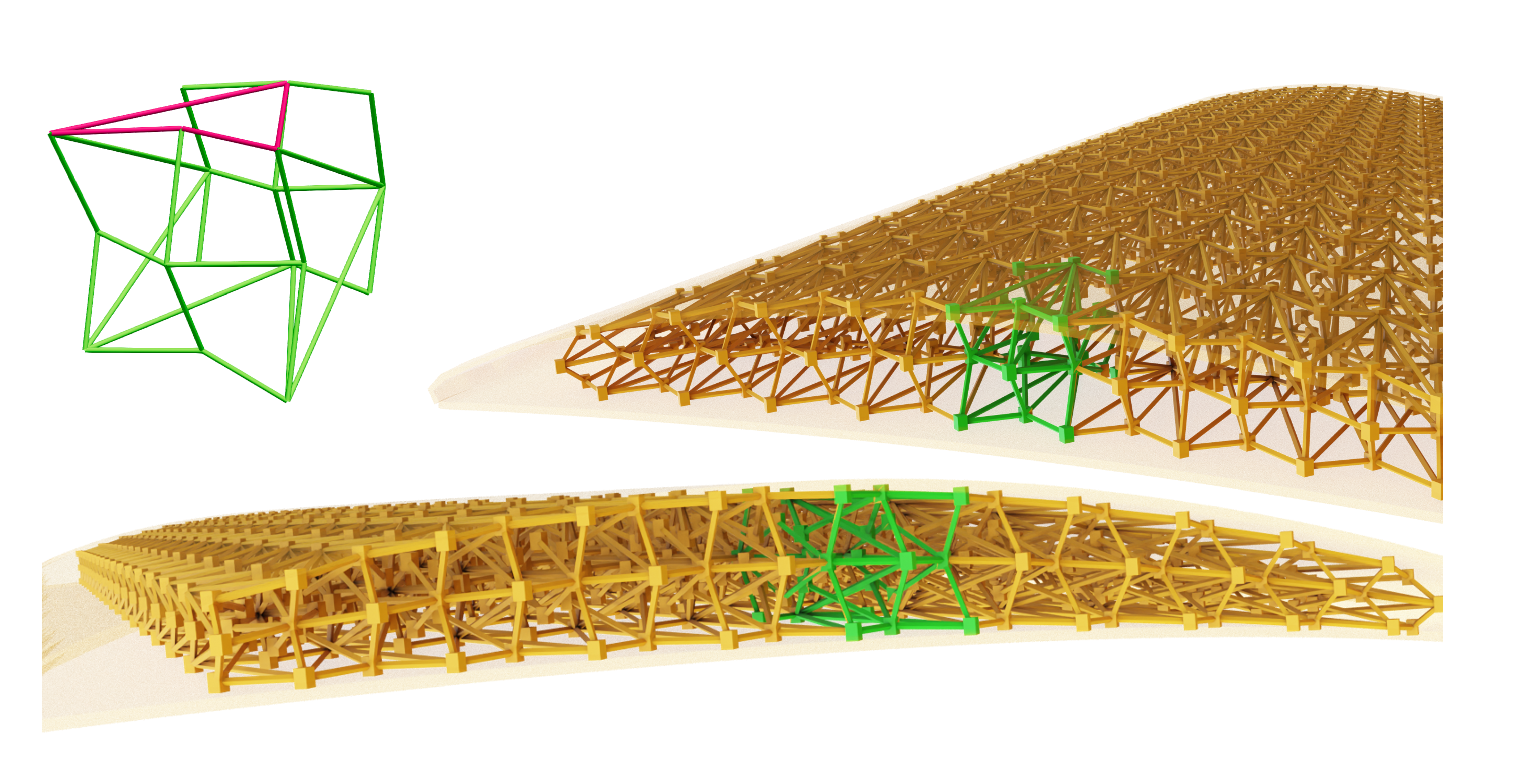}\end{overpic}
\end{center}
\vspace{-15pt}
	\caption{Auxetic 3D double-V structure (top-left). Adapted auxetic structure integrated into the blade cavity seen from two different perspectives.
 }
\label{fig:3DdoubleV}
\end{figure}

Auxetic structures are architectured lattice structures with negative Poisson's ratio \cite{Lim-2015-AuxeticBook}, i.e., they counter-intuitively expand (shrink) transversely under uniaxial longitudinal stretching (compression). An example of a auxetic 2D structure is shown in Fig.~\ref{fig:2DdoubleV}. Compared to conventional materials, auxetic materials exhibit unusual mechanical properties, that have been extensively studied \cite{Greaves2011}. In \cite{chiralWing} an academic 2D prototype airfoil with an auxetic core was studied. In this section, we show how to adapt a 3D auxetic structure to the manufacturability constraints of 3D metal printers. For the blade, we select a suitable microstructure primarily on the basis of manufacturing criteria, as we want to produce a real specimen at the end of our pipeline. The blade with microstructured interior is fabricated layer-by-layer using additive manufacturing, in particular LPBF. 
%process LPBF, see Figure \ref{fig:manufacnalysis:geometry_graded_non_graded}(a).
With additive manufacturing, support structures typically have to be added to the workpiece, and these artificial parts are typically generated by the software that creates the print instructions. These support structures must then be removed after fabrication. One of the easiest ways to reduce the need for support structures is to choose the best orientation for the model to avoid overhangs, bridges, and gaps that require support. However, when manufacturing an object in which the designed microstructure is enclosed inside a closed freeform B-spline shell, there is no way to remove the additional support structures. 
Instead of using software approaches that help mitigate the need for support through simulation and optimized orientation \cite{Vanek2014_supportStructure}, we use a variant of the re-entrant 3D quadrilateral double-V auxetic microstructure \cite{Yang2019_doubleV}, where we connect the hanging nodes by an almost vertical strut composed of two edges connected by a central node, so that the auxetic behavior is preserved, see Fig.~\ref{fig:3DdoubleV} of a standard 3D cell (inset) and an embedding inside a spline surface. 

%This structure does not require any support if it is oriented in the printing direction and the maximum slopes in the vertical (growing) direction is below a manufacturing threshold, see Fig.~\ref{fig:2DdoubleV}(left).
This structure does not require support if it is well aligned in the printing direction and designed so that the maximum angle of the struts relative to the vertical (direction of growth) is below a threshold value for manufacturing, $60^\circ$ in the case of LPBF.
In the case of the blade, the printing orientation is set vertically as shown in %Fig.~\ref{fig:analysis:geometry_graded_non_graded}(a) 
Fig.~\ref{fig:manufacturing:geometry_graded_non_graded}(a). Although one could further optimize the orientation of the blade for 3D printing, e.g., to minimize the support material \cite{Ezair-2015-OrientationOpt}, there are constraints that come from the fact that support material is contained in the outer shell of the blade and must be removed through a few small holes set on the top. We believe that optimization of this kind goes beyond the scope of this paper.
% 
%\mb{I believe that one needs to get out the extra powder}
%When designing the microstructure elements, one has to ensure that all parts of the microstructure satisfy the LPBF threshold of $60^\circ$, which is exactly what our algorithm does. The final auxetic design is shown in Fig.~\ref{fig:3DdoubleV}.

In order to fill a cavity bounded by a freeform shape, represented as a B-spline surface $\Phi^{off}$, recall Fig.~\ref{fig:TestCaseGeo},
%\ref{fig:bladeGeometry}
we take advantage of a relatively simple, cuboidal-like shape of the microstructure tile. Similar to Section~\ref{ssec:StaticMicro}, the cavity is discretized by a hex-mesh with hexahedral cells and the auxetic unit structures are embedded in the hex cells by affine deformations. The periodicity of the structure ensures correct binding in horizontal and vertical directions. 
%Note that the extreme cells of the hex-mesh undergo strong affine deformations where one of the six facets is much smaller than the others. These cells will remain solid for AM. %The auxetic unit cell shown in Fig. 10 has geometric parameters that can be modified, resulting in different angles and lengths of the struts.

%%%%%%%%%%%%%%%%%%%%%%%%%%%%%%%%%%%%%%%%%%%%%%%%%%%%%%%%%%%%%

%\SGPc{TODO: add refs about auxetics for airfoils, etc.}
%%%%%%%%%%%%%%%%%%%%%%%%%%%%%%%%%%%%%%%%%%%%%%%%%%%%%%%%%%%%%
\section{Analysis and microstructural optimization}\label{sec:Analysis}
%%%%%%%%%%%%%%%%%%%%%%%%%%%%%%%%%%%%%%%%%%%%%%%%%%%%%%%%%%%%%

%\input{3_analysis_and_optimization}

Turbine blades need to sustain loads that represent the blades' working conditions, e.g., to be resilient towards centrifugal forces that act on the blades when spinning at high rotation rates. In this section, the turbine blade designs presented in the previous section are analyzed and optimized towards their functionality, demonstrating the seamless interoperability between design and analysis, for both traditional Finite Element Method (FEM) and Isogeometric Analysis (IGA) methods.

\begin{figure}[!tbh]
\hfill
  \begin{overpic}[width=0.9\columnwidth]{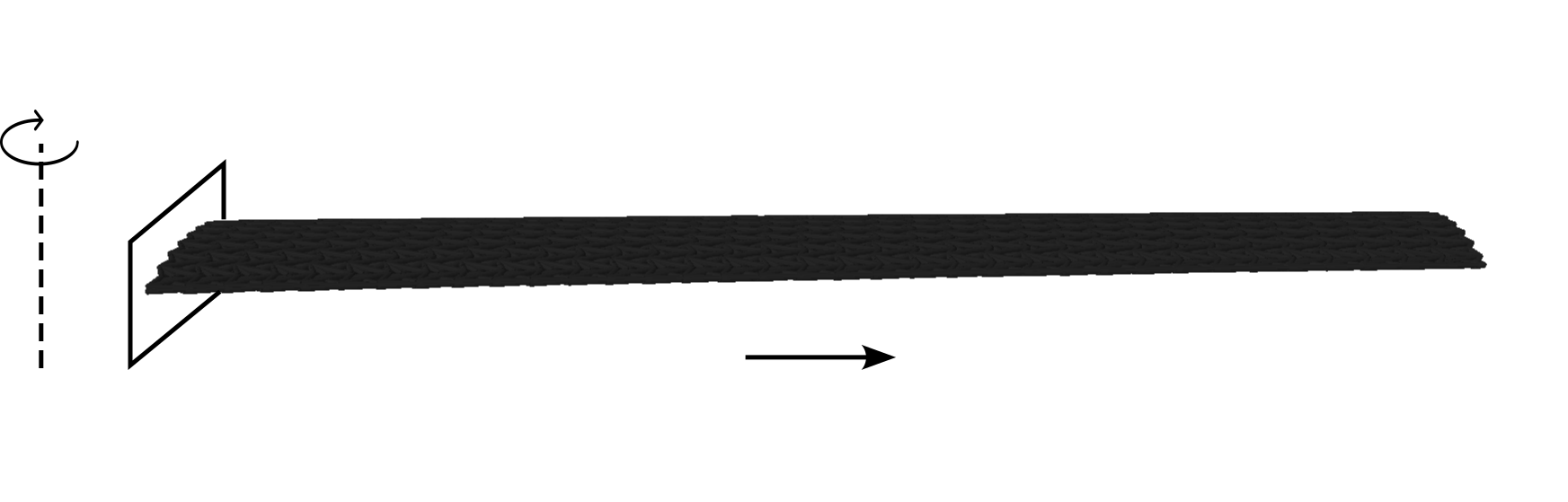}
  \put(18,20){fixed boundary}
  \put(-10,2){axis of rotation}
  \put(43,3){$F_c=m\omega^2r$}
  \end{overpic} 
  \vspace{-8pt}
\caption{Boundary conditions and centrifugal loading of the blade. The centrifugal load $F_c$ is applied on each finite element as a volumetric force, taking into account its mass and the distance $r$ to the axis of rotation (dashed).}\label{fig:analysis:boundary_conditions}
\end{figure}

\begin{figure*}[!tbh]
\begin{center}
\begin{overpic}[width=0.56\textwidth]{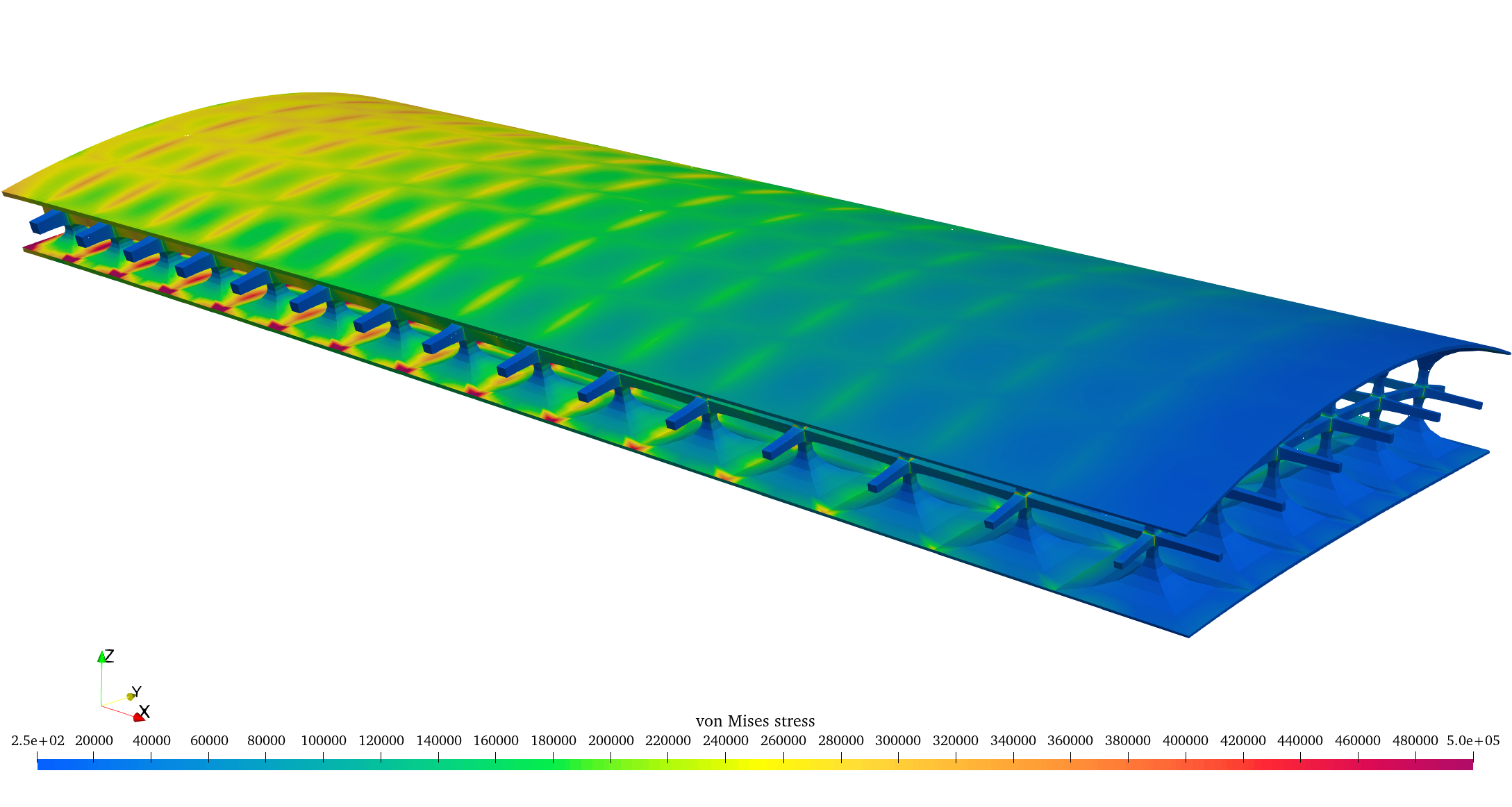}
\put(-5,0){(a)}
\end{overpic}
\begin{overpic}[width=0.34\textwidth]{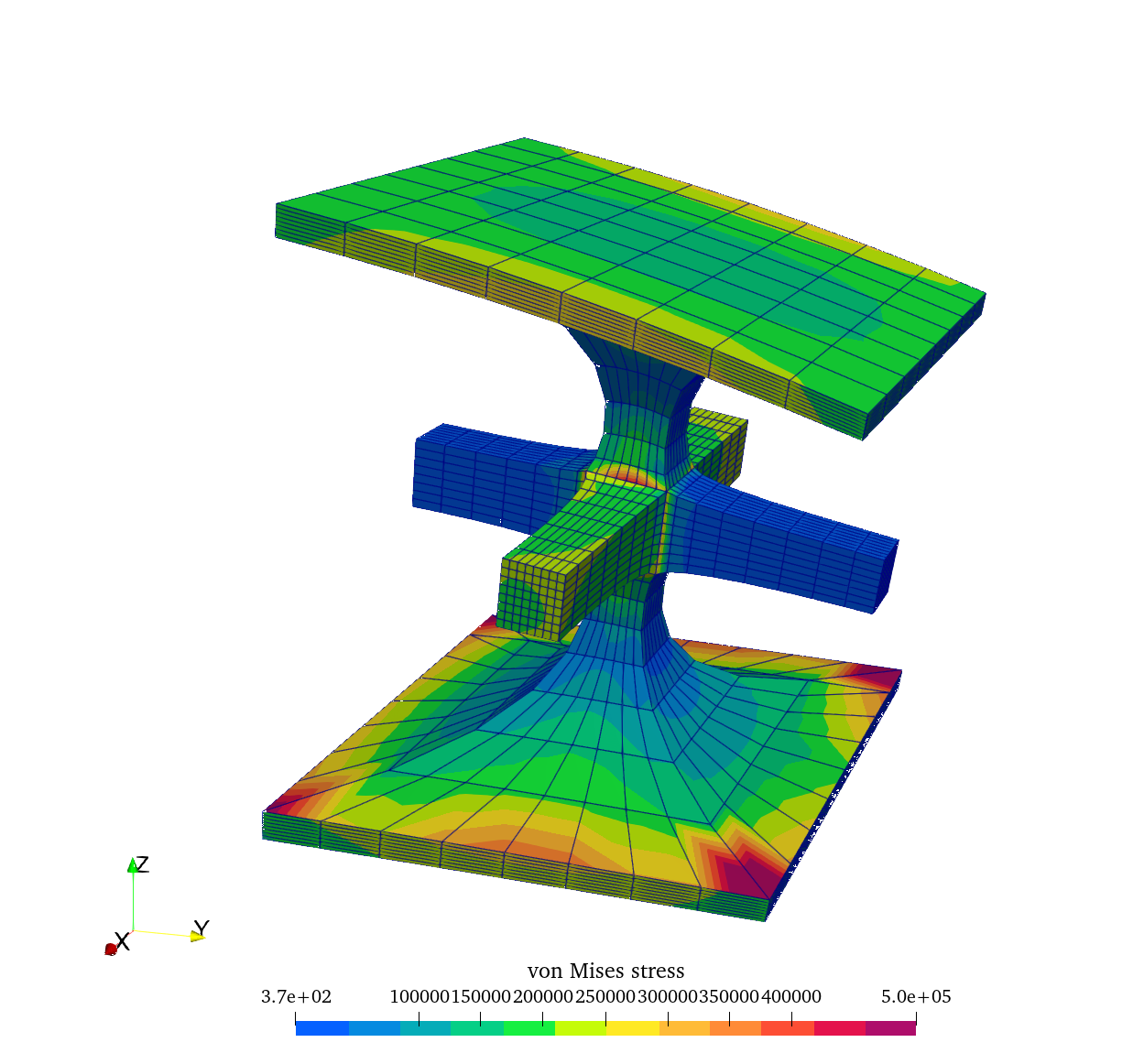}
\put(10,0){(b)}
\end{overpic}
%\begin{picture}(0,0)
%    \put( -475, -3){(a)}
%    \put( -150, -3){(b)}
%\end{picture}
\end{center}
\vspace{-14pt}
\caption{Contourplots of the von Mises stress magnitude in kPa for a graded microstructure using Isogeometric Analysis: (a) stress distribution over the entire blade and (b) detail of a single tile. Note that smallest stress is in the semi-horizontal direction of the tile that goes in the direction of the shorter side of the blade, i.e., the tile in (b) is rotated by $90^\circ$ w.r.t the (a) viewpoint.
}
\label{fig-blade-iga}
\end{figure*}

The blades are subjected to both aerodynamic loads and centrifugal forces caused by their rotation around the central axis of the blisk, the latter being much higher.
Therefore, in this section, we focus on the analysis of a single blade subjected to centrifugal forces, as shown in Fig.~\ref{fig:analysis:boundary_conditions}. A rotational speed of 10,000~rpm is applied and the distance between the fixed boundary of the blade and the axis of rotation is 350~mm. The rotation yields deformations of the blades, which vary with the blade designs. The different resulting deformations are numerically predicted based on simulations of corresponding static structural problems. In detail, linear elasticity model with the material parameters of Inconel 
alloy 718~\cite{inconel718_material}, the material that is used for printing the blades (see later Section~\ref{sec:Manuf}), is assumed.
Therefore the parameters are set as follows:
ndensity of 8.22 g/cm$^3$, Young's modulus of 208 GPa, and Poisson's Ratio of 0.3.
In addition, the blade is considered to be fully attached to the disk, so all displacement degrees of freedom are set to zero at the blade section in contact with the disk. This is in accordance with engineering practice; even though the blades are sometimes manufactured separately, they are mounted to a disk using a slide-in mechanism that leaves no degree of freedom to the blade to move with respect to the central disk.

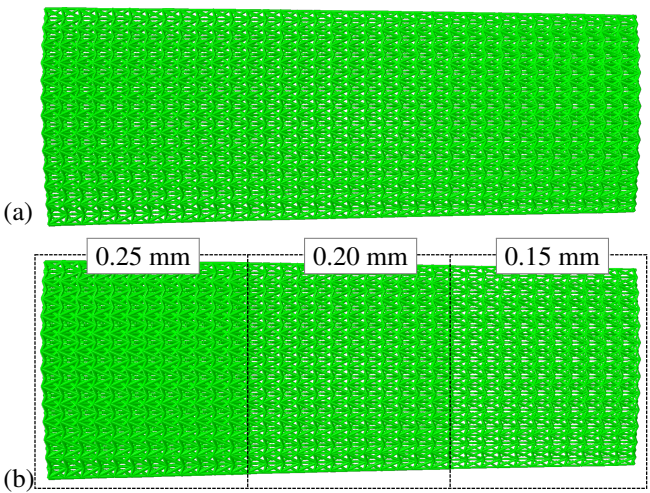
\begin{figure}[!tbh]
\hfill
  \begin{overpic}[width=0.95\columnwidth]{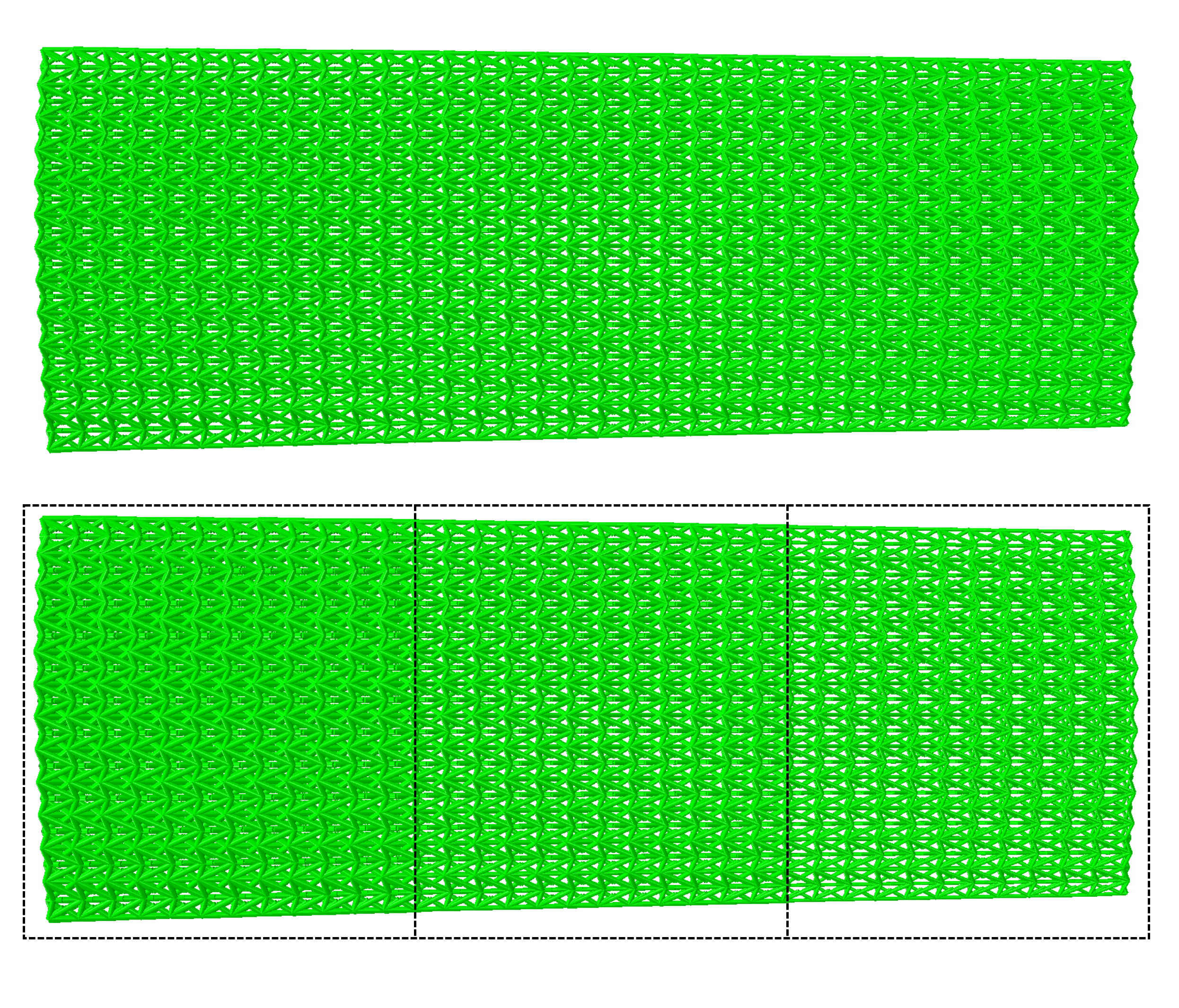}
  \put(10,41){\fcolorbox{gray}{white}{0.25~mm}}
  \put(44,41){\fcolorbox{gray}{white}{0.20~mm}}	
  \put(74,41){\fcolorbox{gray}{white}{0.15~mm}}	
  \put(-3,48){(a)}
  \put(-3,6){(b)}
  \end{overpic} 
\hfill \vrule width0pt\\
\vspace{-30pt}
\caption{Auxetic geometries used in the simulation. (a) A uniform auxetic structure consisting of tiles with constant thickness of 0.2~mm. (b) A graded version of the same microstructure with three distinct regions of radii (dashed black), the thickest tiles being used closer to the base of the blade (left). %0.25~mm, 0.2~mm, 0.15~mm respectively.
}\label{fig:analysis:geometry_graded_non_graded}
\end{figure}

%%%%%%%%%%%%%%%%%%%%%%%%%%%%%%%%%%%%%%%%%%%%%%%%%%%%%%%%%%%%%
\subsection{Analysis of Microstructures}
%%%%%%%%%%%%%%%%%%%%%%%%%%%%%%%%%%%%%%%%%%%%%%%%%%%%%%%%%%%%%

This section highlights the suitability of the proposed workflow in practical engineering contexts, by demonstrating the compatibility with commercial FEM software, specifically Abaqus, and IGA simulation tools.
In this example, an auxetic microstructured geometry, shown in 
Fig.~\ref{fig:analysis:geometry_graded_non_graded},
% \mb{Pablo, this Fig. seems broken, but I would suggest to include the miscostructure in your software. Or cannot we refere to the current Fig. 11 (Contourplots of displacement)?}
is analyzed.

\begin{figure}[!tbh]
\hfill
  \begin{overpic}[height=0.25\columnwidth]{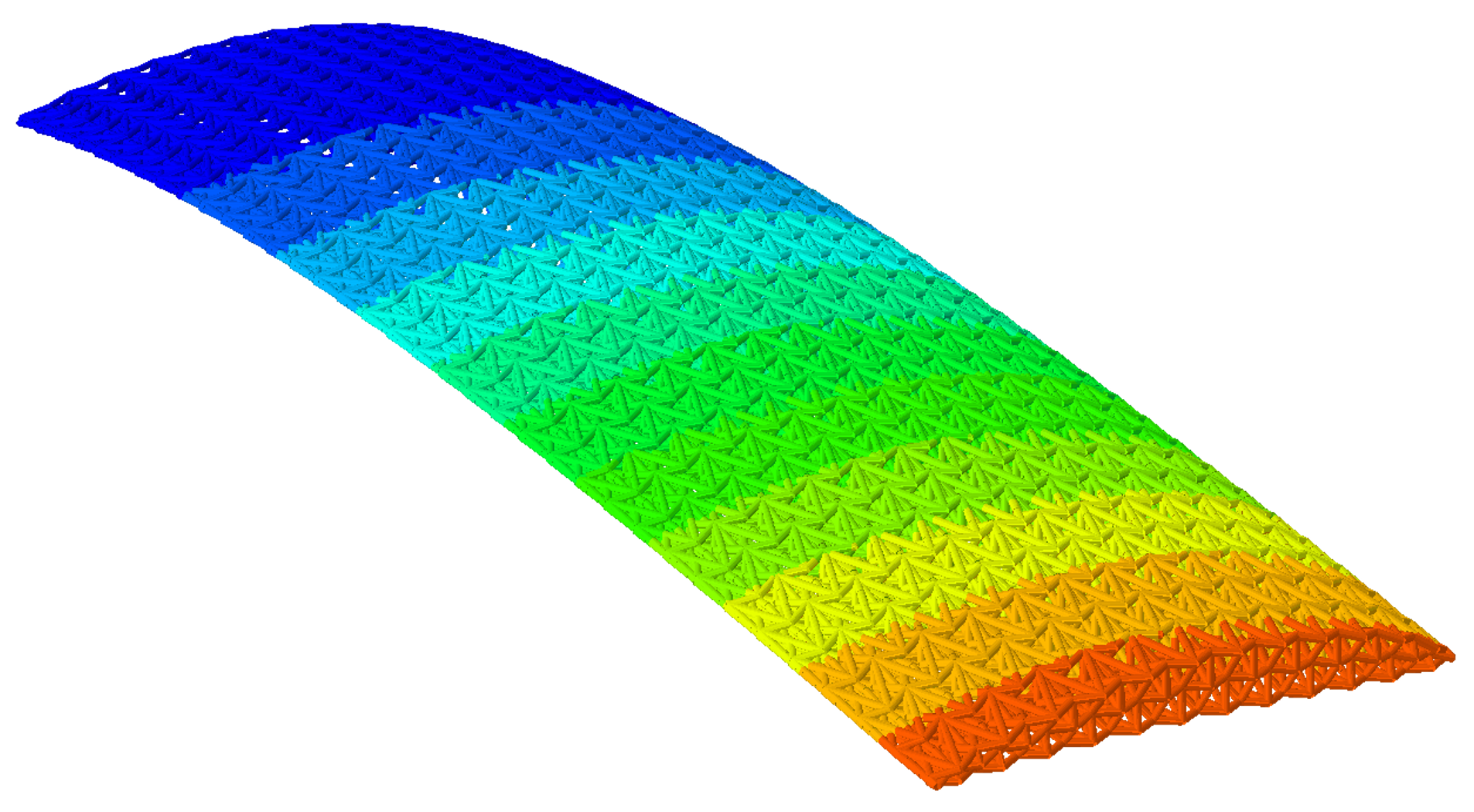}
  \put(25,15){(a)}
  \end{overpic} 
  \hfill
  \begin{overpic}[height=0.25\columnwidth]{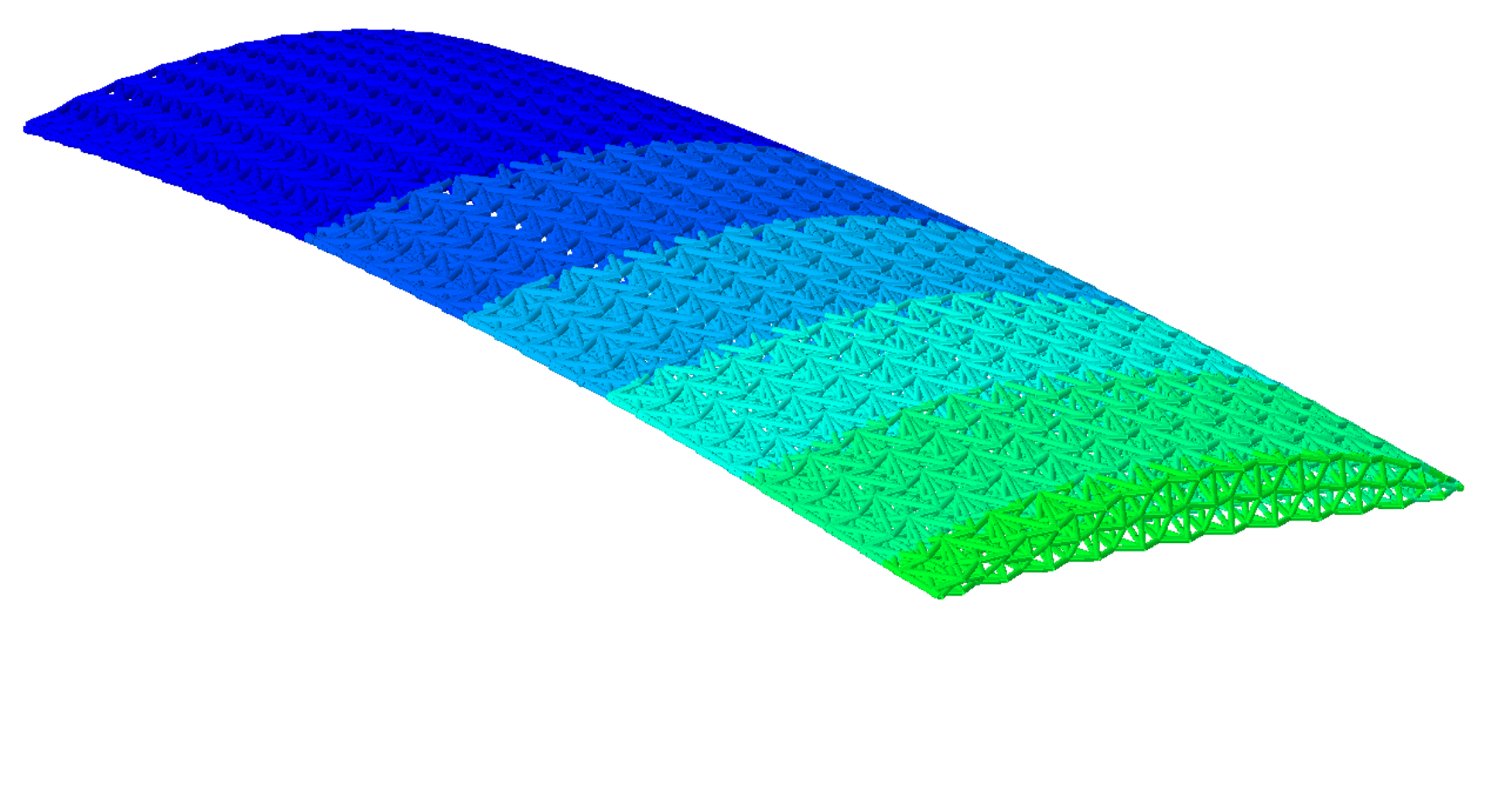}
  \put(35,15){(b)}
    \put(-70,52){\fcolorbox{gray}{white}{fixed base of the blade}}	
    \put(-5,-12){\includegraphics[width=0.15\columnwidth]{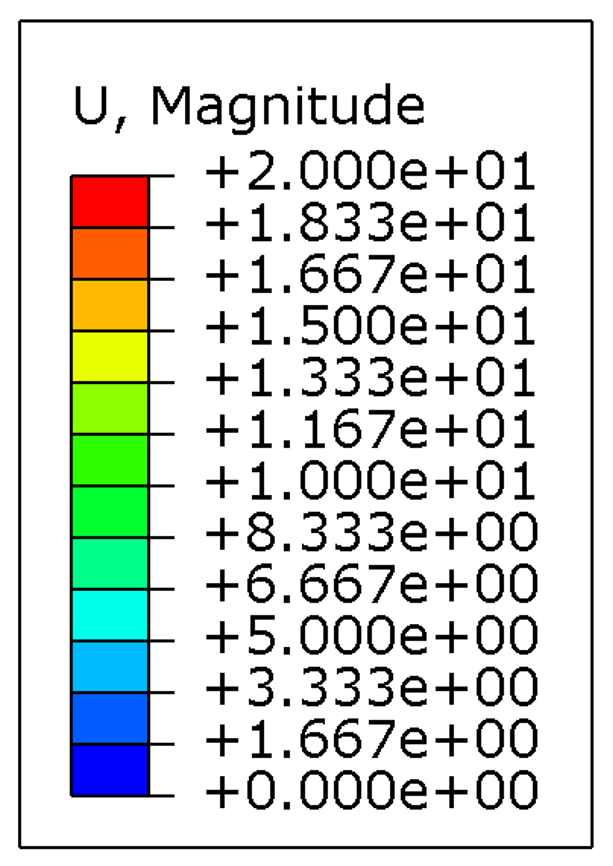}}
  \end{overpic} 
\hfill \vrule width0pt\\
\vspace{-6pt}
\caption{Contourplots of the displacement magnitude in mm for the two auxetic microstructures (recall Fig.~\ref{fig:analysis:geometry_graded_non_graded}): (a) While the uniform microstructure of constant thickness of 0.2 mm undergoes a deformation of up to 20 mm on the free end, (b) the graded counterpart, whose tiles are thicker closer to the fixed base and thinner on the other end, is displaced considerably less. 
}\label{fig:analysis:results_graded_non_graded}
\end{figure}

In the case of FEM, two configurations are examined: a uniform microstructure with constant strut radii of 0.2~mm, Fig.~\ref{fig:analysis:geometry_graded_non_graded}(a), and a graded version of the same microstructure with three distinct regions of radii 0.25~mm, 0.2~mm, 0.15~mm, Fig.~\ref{fig:analysis:geometry_graded_non_graded}(b).
This choice is motivated by the centrifugal loading that acts on the turbine blades, which increases linearly with the distance to the center of rotation.

Since the geometry consists of slender struts, it is reasonable to perform finite element analyses utilizing beam elements for efficient calculations. In particular, the Abaqus element type B31, which uses Timoshenko beam theory~\cite{timoshenkoStrengthMaterials1956}, is chosen.
A total of 24101 elements and 16294 nodes are used to discretize the blade geometry. 
Fig.~\ref{fig:analysis:results_graded_non_graded} shows the numerically predicted deformation of the structure, where (a) shows the results of the uniform microstructure and (b) the results of the graded version.
The maximum deflection of the graded version is decreased by a factor of two when compared to the uniform design.

In order to demonstrate the interoperability between analysis suitable designs (see Section~\ref{sec:Design}) and IGA simulation tools, we analyzed a graded blade tile directly using the spline description of the design. The considered tile design is the one in Fig.~\ref{fig-blade-lattice-tiles}(a), while the blade macro-shape is the same as in previous cases.
Fig.~\ref{fig-blade-iga} shows the calculated stress distribution applying same material and loading conditions as before.
In Fig.~\ref{fig-blade-iga}(b), a refined version of the tile, preserving the original parameterization, was applied to each individual tile in order to accurately capture its elastic behavior.
% \mb{Figure~\ref{fig-blade-iga} is great. Pls comment on the (b) image, the blue (small stress) direction vs. the transversal (greenish) direction. This something that one expects...and is nicely seen from the simulation.
The simulation was performed with the IGA solver YETI~\cite{yeti}.
Fig.~\ref{fig-blade-iga}(a) shows how the higher stress concentration observed near the root of the blade is accounted for using thicker tiles, while a thinner design is used near the tip.
In addition, the stress distribution shows a clear anisotropy due to the application of centrifugal loads along the $x$-axis: As can be seen in the four arms of the tile in Fig.~\ref{fig-blade-iga}(b), the longitudinal stresses ($x$-axis) are higher than the transverse stresses ($y$-axis).
The model in Fig.~\ref{fig-blade-iga}.
The simulation, which includes 99'819 degrees-of-freedom, was executed on a single thread of an Apple M2 Max processor. The computation itself required 8.85 seconds and approximately 2 GB of memory. Generating the output files took an additional 6.4 seconds.

%
%\begin{figure}[!tbh]
%\hfill
%  \begin{overpic}[width=\columnwidth]{fig/analysis_auxetic_structure_comparison_graded_non_graded_no_caption.png}
%  \put(30,60){(a)}
%  \put(30,20){(b)}
%  \end{overpic} 
%\hfill \vrule width0pt\\
%\vspace{-60pt}
%\caption{Contourplots of the displacement magnitude in mm for the auxetic microstructures: (a) results of the uniform microstructure and (b) results of the graded version. 
%}\label{fig:analysis:results_graded_non_graded}
%\end{figure}

\begin{figure}[!tbh]
  \centering
  \includegraphics[width=.95\columnwidth]{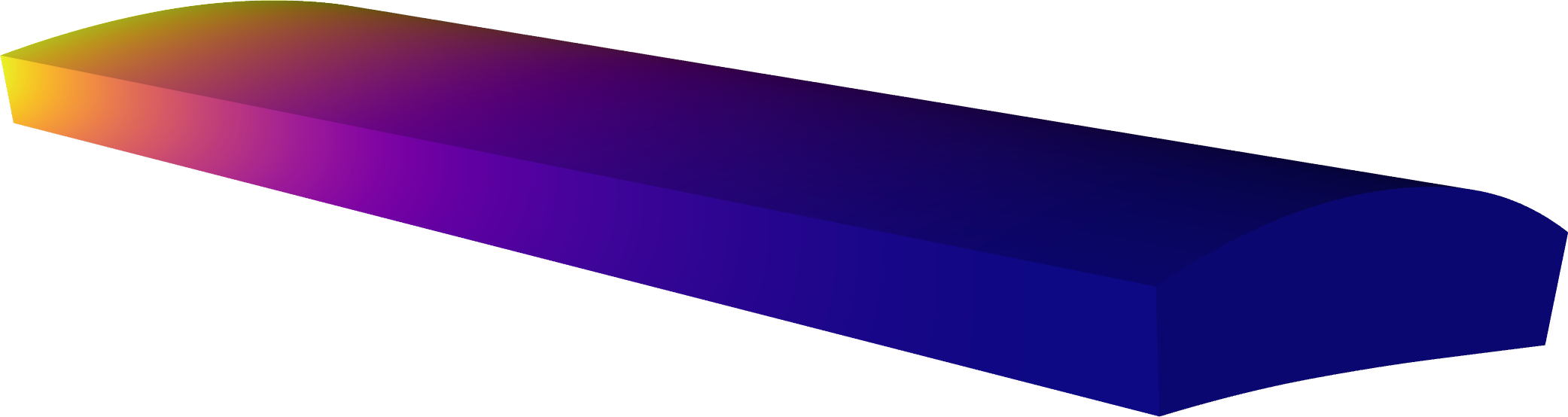}
  \put(-240,25){(a)}
  \\
  \includegraphics[width=.95\columnwidth]{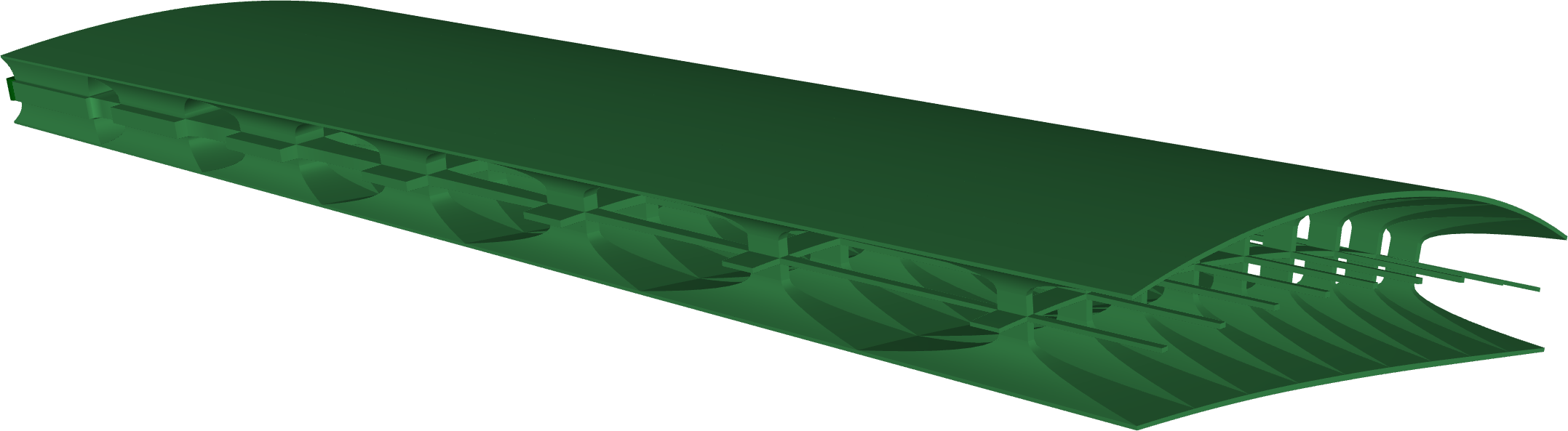}
  \put(-240,25){(b)}
  \caption{
    The automatically optimized thicknesses of the tiles' arms (a), which are thick (yellow) close to the axis of rotation and thin (blue) close to the blade's tip, as well as the corresponding optimized microstructured blade (b).}
  \label{fig:designOptimized}
\end{figure}

\subsection{Optimization of Non-Homogeneous Free-Form Microstructures}\label{ssec:OptNonHom}

As non-homogeneous microstructures can increase the resistance of the blade towards the centrifugal force (see above), the question arises whether this resistance can be even further increased by an automatic optimization of non-homogeneous free-form microstructures.
Thus, an automatic optimization framework that builds on
\begin{enumerate}
  \item Splinepy \cite{splinepy} for the creation of the computational geometries based on the design parameters,
  \item G+Smo \cite{gismo} for the prediction of deflections and their sensitivities with respect to the design parameters,
  \item as well as SciPy \cite{2020SciPy-NMeth} for the gradient-based optimization of the blades' compliances with respect to the design parameters
\end{enumerate}
was developed.
% To simplify the argument, we do not consider a free-form outer shape; instead the blades' shape is simplified as a rectangular box with a width of 34~mm, a height of 9~mm, and a length of 89~mm.
Similarly to Section \ref{ssec:StaticMicro}, the free-form outer shape is used to create the microstructure, with the difference that no lower-order spline approximation is performed.
The free-form microstructure is based on tiles that resemble the one depicted in Fig.~\ref{fig-blade-lattice-tiles}(a), i.e., the cross-like tile having axis-parallel arms with parameterized thicknesses.
The individual thicknesses of the arms are determined by evaluating a so-called parameter spline~\cite{Zwar2022}, which is in this case chosen to be a quadratic B-spline.
The coefficients of this parameter spline are the design variables that are optimized automatically.
Initially, the coefficients are set to a value that mimics the solid blade design.

After nine gradient-based optimization steps (Sequential Least Squares Programming), the automatically optimized design resembles the manually optimized design: it is almost solid close to the rotation axis and the microstructure becomes thinner towards the blade's tip, i.e., with increasing distance to the rotation axis, see Fig.~\ref{fig:designOptimized}.
Compared to the (almost) solid blade design, this optimally microstructured design achieves a weight reduction higher than 50\% and, at the same time, a compliance reduction of more than 75\%.

% \begin{figure}[!tbh]
% \hfill
%   \begin{overpic}[angle=0,width=0.48\columnwidth]{fig/designInitial.png}
%   \put(0,65){\fcolorbox{gray}{white}{\includegraphics[angle=0,,width=0.11\textwidth]{fig/tileInitial.png}}}  
%   \put(-2,2){(a)}
%   \end{overpic}
% \hfill
%   \begin{overpic}[angle=0,width=0.48\columnwidth]{fig/designOptimal.png}
%    \put(0,65){\fcolorbox{gray}{white}{\includegraphics[angle=0,,width=0.11\textwidth]{fig/tileOptimal.png}}}  
%   \put(-2,2){(b)}
%   \end{overpic} 
% \hfill \vrule width0pt\\
% \vspace{-15pt}
% 	\caption{Lower halves of the initial (left) and optimized (right) blade designs. The framed boxes show close-ups of the selected tile. }\label{fig:designOptimized}
% \end{figure}

%%%%%%%%%%%%%%%%%%%%%%%%%%%%%%%%%%%%%%%%%%%%%%%%%%%%%%%%%%%%%
\section{Hybrid manufacturing}\label{sec:Manuf}
%%%%%%%%%%%%%%%%%%%%%%%%%%%%%%%%%%%%%%%%%%%%%%%%%%%%%%%%%%%%%

To manufacture a metallic blade with internal microstructures, we opted for a laser powder bed fusion (LPBF), followed by 5-axis CNC machining. While LPBF was used to manufacture the shell and the interior microstructures, 5-axis CNC machining was used to achieve a high-quality surface finish. 

% \begin{figure}[!tb]
% \hfill
%   \begin{overpic}[width=0.88\columnwidth]{fig/2.png}
%   \end{overpic} 
% \hfill \vrule width0pt\\
% \vspace{-16pt}
% \caption{(a) Five specimens of the microstructured blades after PBF-LB. b) Accelerometer positions and excitation points
% .}\label{fig:manufacturing:geometry_graded_non_graded}
% \end{figure}

% \The 
The blades were produced using a Renishaw® AM400 machine equipped with an Yttrium fiber laser capable of delivering a maximum power of 400 W. Inconel 718® gas-atomized powder was used as the material for this manufacturing process. The particle size distribution was assessed in accordance with ISO 13320 standards. The cumulative percentages for the gas-atomized powder were as follows: d1 = 10.0 µm, d10 = 17.1 µm, d50 = 32.5 µm, and d90 = 54.8 µm, indicating the size distribution at specified volume percentages.

\begin{figure}[!tb]
\begin{center}
\hfill
\begin{overpic}[height=0.42\columnwidth]{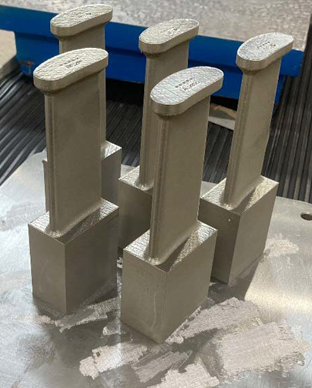}
\put(-11,0){(a)}
\end{overpic} 
\hfill
\begin{overpic}[height=0.42\columnwidth]{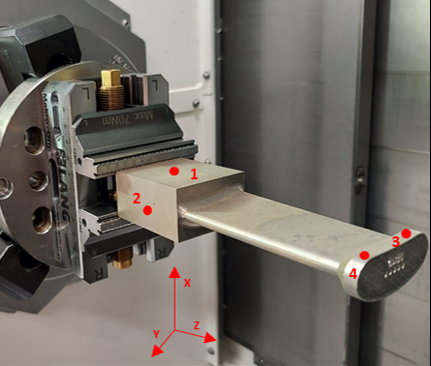}
\put(-11,0){(b)}
\end{overpic} 
\hfill \vrule width0pt\\
\end{center}
\vspace{-16pt}
\caption{(a) Five specimens of the microstructured blades after PBF-LB. (b) Accelerometer positions and excitation points
.}\label{fig:manufacturing:geometry_graded_non_graded}
\end{figure}

The Powder Bed Fusion with Lattice Blades (PBF-LB) manufacturing strategy employed a raster technique with a 67° rotation between layers (10 mm stripes). 
The blade is 90 mm long and cca 40 mm wide, mounted on a cuboidal base of $29\times48\times 54$ mm. The blade thickness varies between 6 and 7.5 mm, largest being in the central area.
To minimize deviations arising from residual stresses inherent in the process, the blades were printed together in a vertical orientation, 
see Fig.~\ref{fig:manufacturing:geometry_graded_non_graded}.
Additionally, a 1.5 mm base stock was added to prevent any geometric issues due to the interface between the pieces and the plate, ensuring dimensional accuracy after machining. The manufacturing parameters used were as follows: a laser power of 200 W, a scanning speed of 1000 mm/s, a hatching distance of 0.09 mm, a layer thickness of 60 µm, and a laser beam size of 70 µm. The smallest cell/unit that one can print using LPBF is around 100 µm (bars diameter), considering that powder spheres diameter is around 20-60 µm. Five microstructured blades were manufactured via the PBF-LB manufacturing process, see
Fig.~\ref{fig:manufacturing:geometry_graded_non_graded}(a), using five different interior microstructures; recall Figs.~\ref{fig:MS12DR} and~\ref{fig-blade-lattice-diagonal} for the original designs we used.

All the produced blades underwent machining on a 5-axis Mazak INTEGREX® i200 smooth multitasking machine. These components were securely held in place using a 5-axis Makro-Grip Lang 48120-77 (clamp chuck with a clamping force of up to 14,000 N) during milling operations, without the use of a tailstock.

\begin{figure*}[!tbh]
\vrule width0pt\hfill
\begin{overpic}[width=0.3\textwidth]{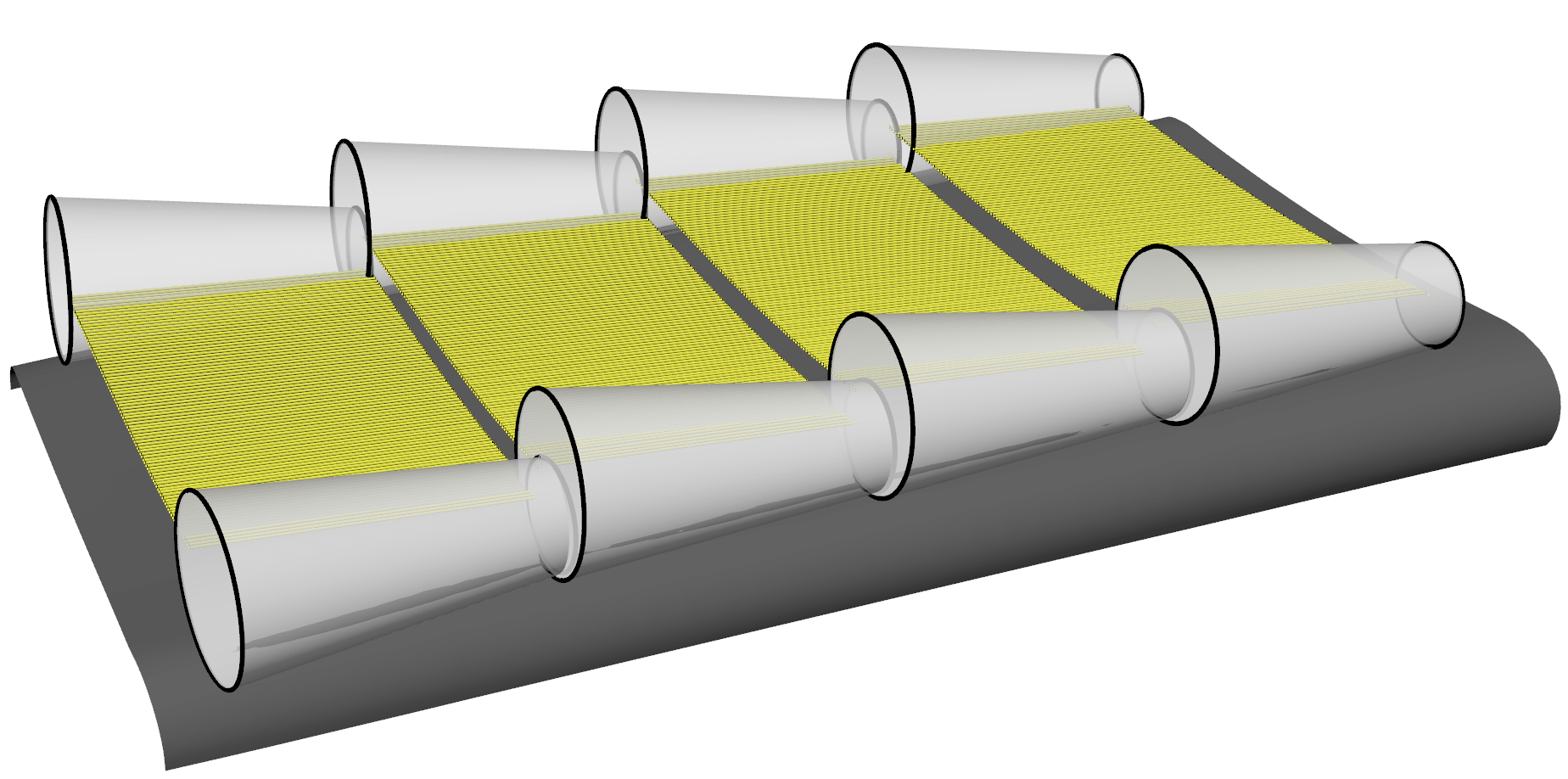}
% \put(50,1){\includegraphics[angle=0,width=0.55in]{fig/colorT}}
%\put(75,1.5){\fcolorbox{gray}{white}{\includegraphics[width=0.1\textwidth]{fig/blade2}}}
%\put(39,0){0.04}
%\put(71,0){-0.04}
\put(-1,-2){(a)}
\put(90,12){$\Phi$}
\end{overpic} 
\hfill 
\vrule width0pt \hfill
\begin{overpic}[width=0.28\textwidth]{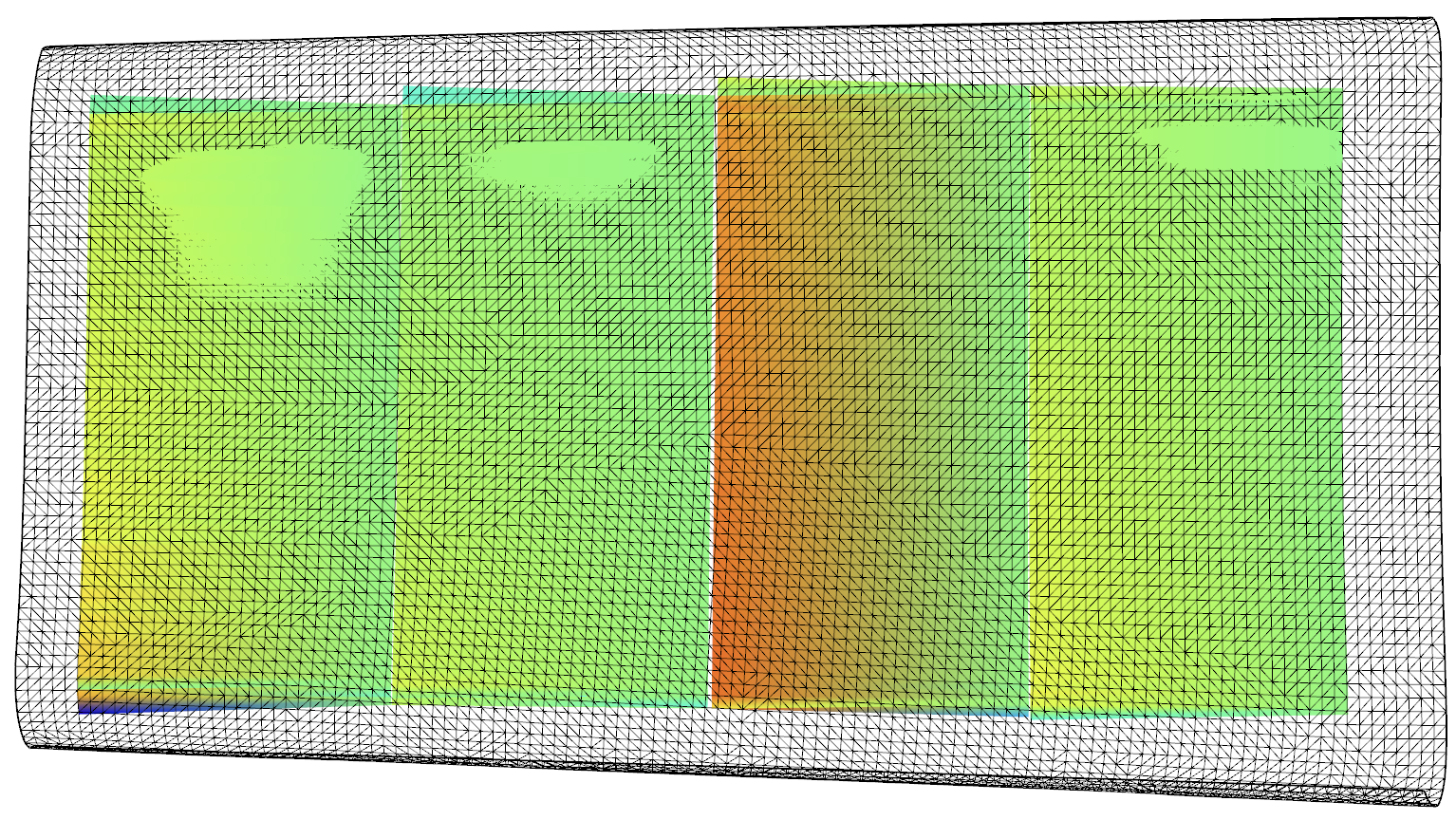}
\put(-5,-2){(b)}
% \put(50,-3){\includegraphics[angle=0,width=0.5in]{fig/tool}}
\end{overpic}
\hfill
\begin{overpic}[angle=90,width=0.012\textwidth]{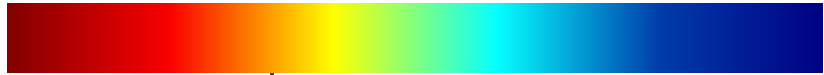}
{\footnotesize
\put(-3,102){0.04}
\put(-5,-8){-0.04}
}
\end{overpic}
\hfill
\begin{overpic}[width=0.28\textwidth]{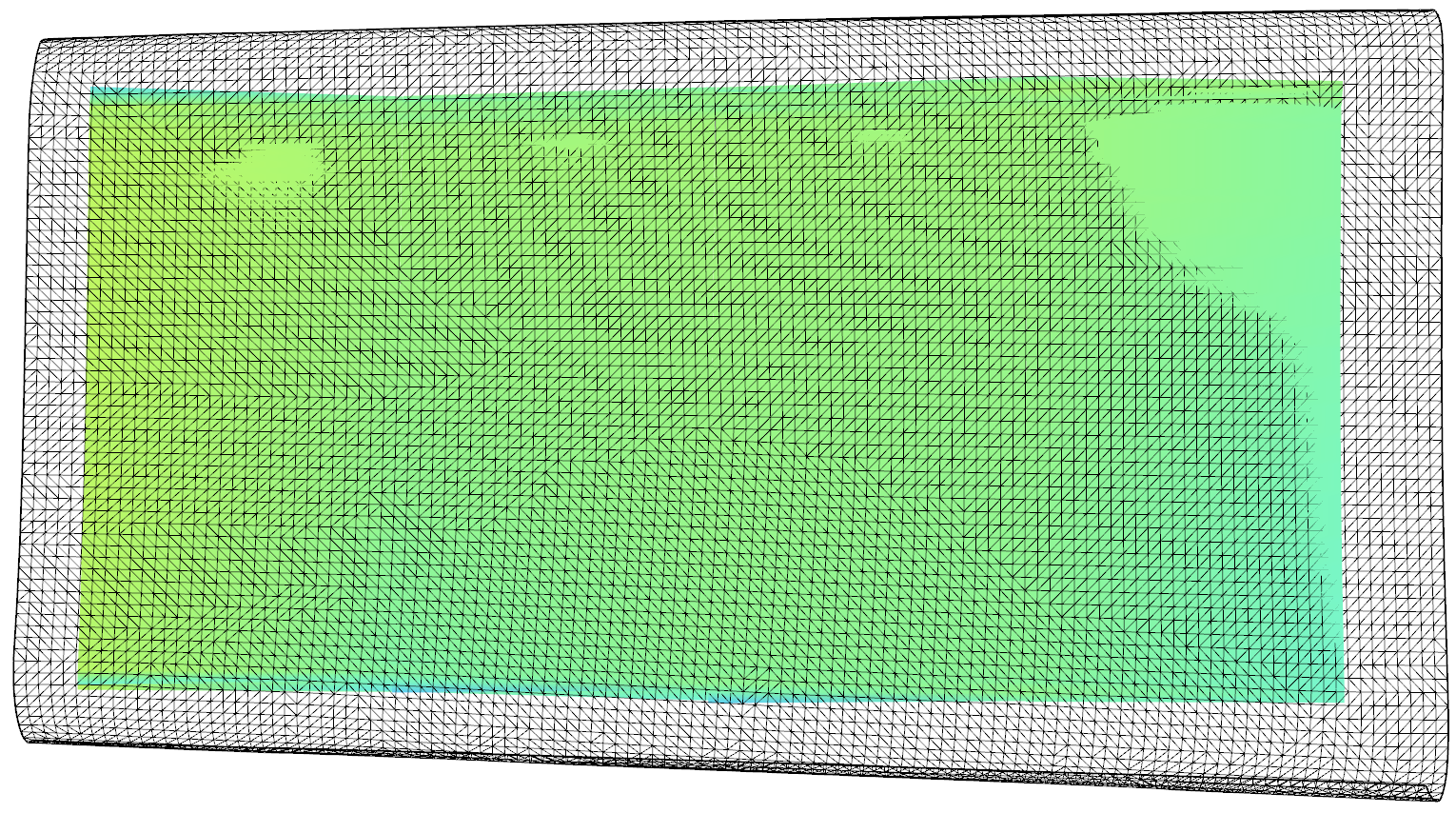}
\put(0,-2){(c)}
\end{overpic}
\hfill \vrule width0pt\\
\vspace{-16pt}
\caption{(a) The bottom side of the boundary of the blade,  $\Phi$, is approximated by four paths of a givel conical tool; only the engaging part of the tool is shown (light grey) together with the motions of the tool's axis, the ruled surfaces (yellow). (b) The envelopes on $\Phi$ are color-coded by the distance error, with error threshold being $\pm$40µm. While the initial paths have visible gaps in between the envelopes strips and posses non-smooth behavior of the distance error, the optimized results towards $G^1$ continuity have smooth behavior of the distance error (c).}\label{fig:blade2}
\end{figure*}

{\textit{Simulation of surface finish via 5-axis flank CNC machining.}
We also experimented with simulations of high-quality surface finish using 5-axis flank CNC using conical tools, see Fig.~\ref{fig:blade2}. We modeled four paths using a given conical cutting tool and optimized the envelopes towards $G^1$-ness between the neighboring envelopes via \cite{Rajain-2023-MultiStripG1}. The optimization stage included a trade-off between approximation quality and $G^1$ continuity. Equal weights were assigned to the $F_{approx}$ and $F_{G^1}$, see~\cite[Section 6, Eq.~(18)]{Rajain-2023-MultiStripG1}, to balance approximation quality and smoothness between neighbouring patches. $F_{fair}$ and $F_{ini}$ were used as regularizers with small weights ($w_4 = w_5 = 0.0001$). The tool axis has to stay rigid throughout the tool trajectory, therefore $F_{rigid}$ was assigned a large weight ($w_1 = 1$). Even though the simulation results were below the given threshold of 50 µm, only two paths would be physically possible due to global collision therefore the surface finish was accomplished using point milling, which we describe next.

\begin{figure}[!tbh]
\hfill
  \begin{overpic}[width=0.88\columnwidth]{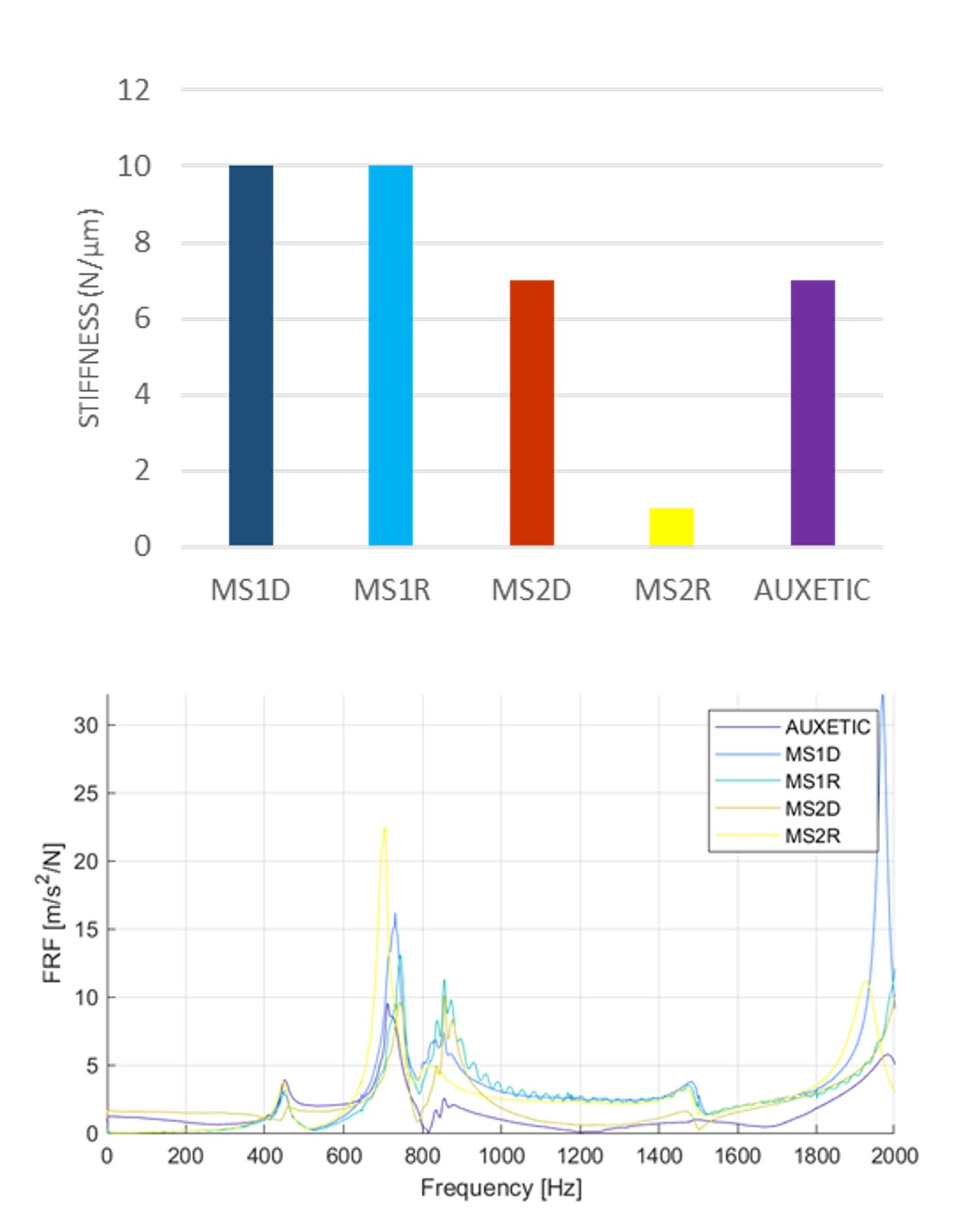}
  \put(0,92){(a)}
  \put(0,40){(b)}
  \end{overpic} 
\hfill \vrule width0pt\\
\vspace{-24pt}
	\caption{Hammer testing for no-machined blades, i.e., the blades after only PBF-LB manufacturing process (accelerometer positioned in position 4 with hammer excitation in position 2), recall 
    Fig.~\ref {fig:manufacturing:geometry_graded_non_graded}(b).  (a) Measured static stiffness. (b) Measured receptance curves.}\label{fig:NoMachined}
\end{figure}

\begin{figure*}[!tb]
\begin{center}
\hfill
\begin{overpic}[height=0.36\columnwidth]{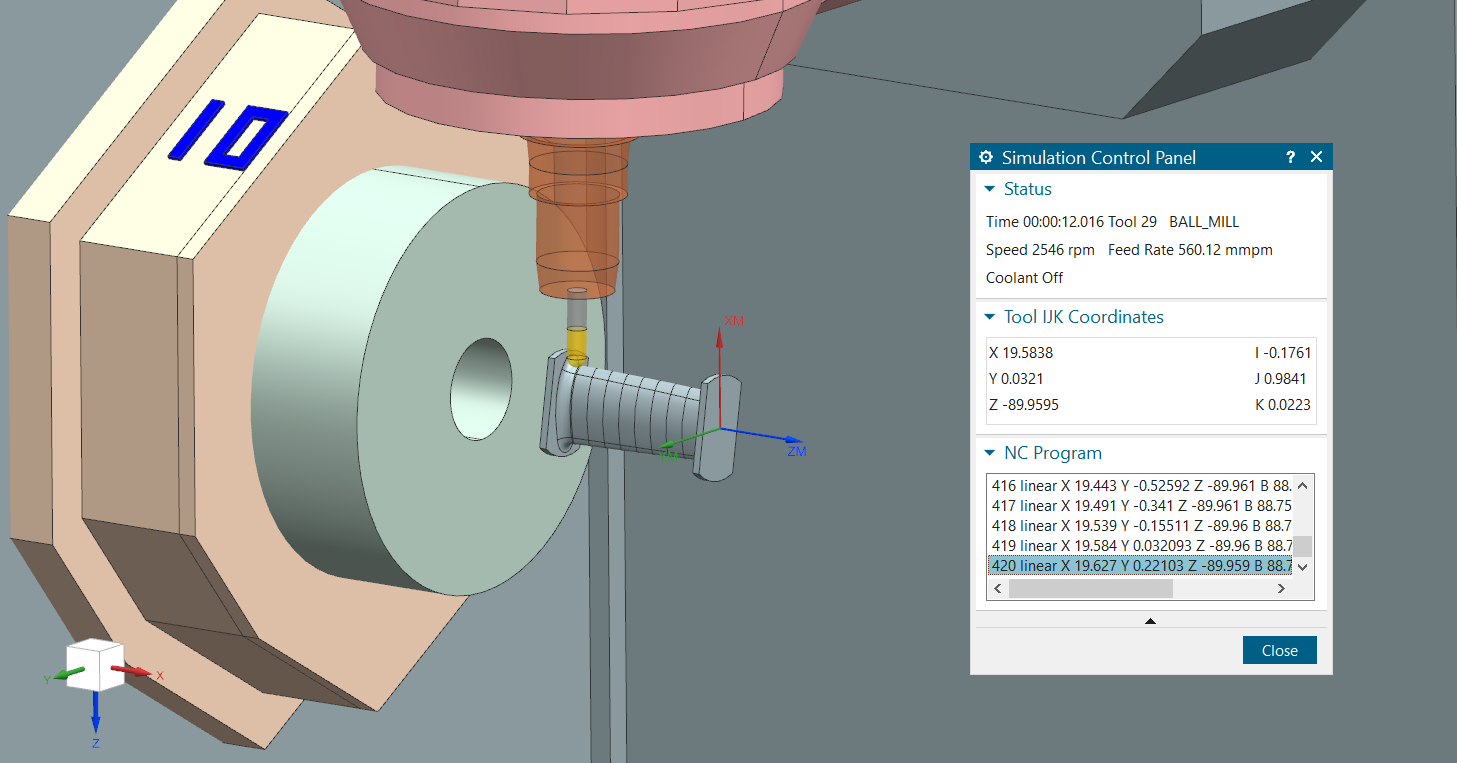}
\put(-11,0){(a)}
\end{overpic} 
\hfill
\begin{overpic}[height=0.36\columnwidth]{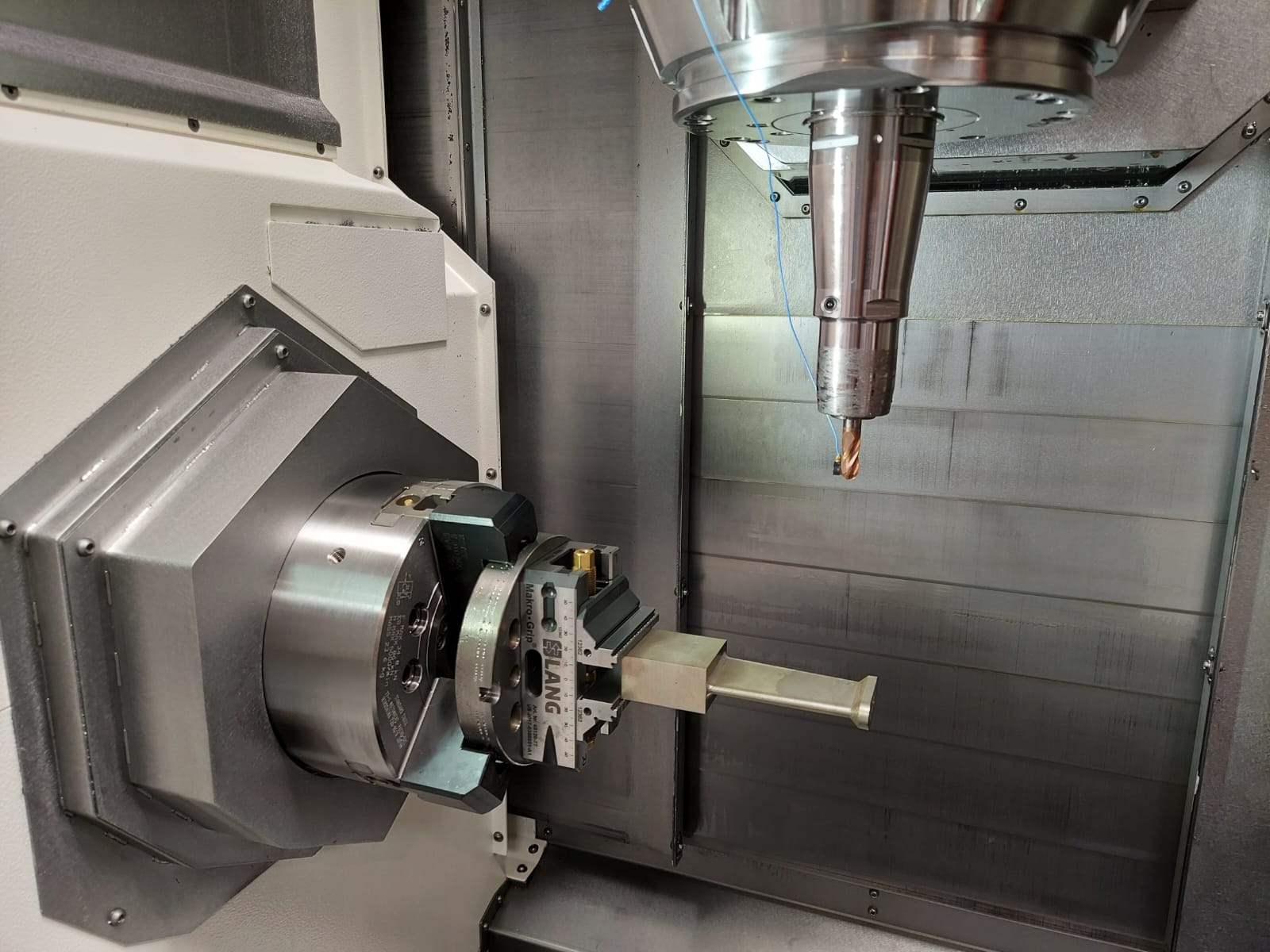}
\put(-11,0){(b)}
\end{overpic} 
\hfill
\begin{overpic}[height=0.36\columnwidth]{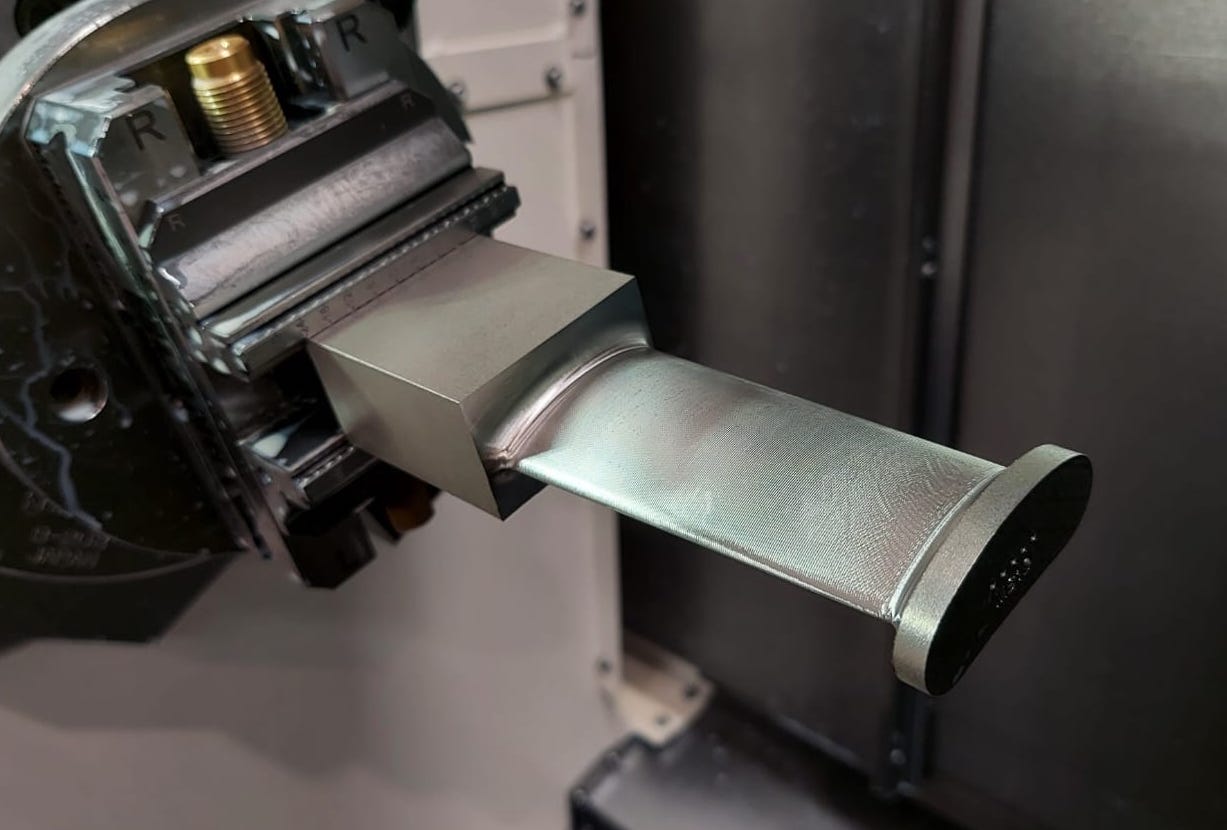}
\put(-11,0){(c)}
\end{overpic} 
\hfill \vrule width0pt\\
\end{center}
\vspace{-16pt}
\caption{Surface finishing using 5-axis CNC machining with ball-end cutter. (a) Simulation using a CAM software, Siemens NX. (b) Workpiece being CNC machined in Mazak INTEGREX® i200  multitasking machine. (c) A zoom-in image of the back side of the blade.}\label{fig:SimulationAndMachining}
\end{figure*}

{\textit{Physical finishing using 5-axis ball CNC machining.}}
%Machining strategies using ball cutter were as follows. Finishing operations were reproduced with different offsets of the final surface of 1.4-0.9-0.45-0 mm until reaching the desired surface. 
The machining conditions were 2546 rpm, 560 mm/flute and 0.4mm radial pass distance. The orientation angle of the tool was 30 degrees for lead angles and zero degrees for tilt angle. The tool was a ball-end milling tool with 10 mm diameter, 4 flutes, helix angle of 30 degrees, and 70 mm length. Cutting tool material is coated with hard metal (HM MG10). Milling tools were mounted in a cylindrical-collect tool holder with about 40 mm cantilever, runout below 5 µm. Tool static run-out and length were verified using a Zoller® SmarTcheck 600 tool presetter. The simulation and physical CNC machining using a ball-end cutter is shown in 
Fig.~\ref{fig:SimulationAndMachining}.

%1.4-0.9-0.45-0 mm

{\textit{Dynamic behaviour analysis.}} 
A hammer testing was conducted, before (Fig.~\ref {fig:NoMachined}) and after (Fig.~\ref {fig:Machined}) CNC machining, for the five microstructured geometries to consider the effects of the machine and clamping systems on the harmonic response of the blades, obtaining the FRF (Frequency Response Function). Thus, two monoaxial accelerometers (PCB Piezotronics 352C22) and a hammer (PCB Piezotronics 3425 with load sensor PCB 0865C03) were used. Excitation points are 1 and 2 and accelerometer positions are points 3 and 4, see Fig.~\ref{fig:manufacturing:geometry_graded_non_graded}(b).

We observed that the only significant results of dynamic behavior were obtained for the accelerometer positioned in 4 with hammer excitation in 2.  Hence, structural properties of the five microstructured blades were experimentally tested, obtaining static stiffness, internal damping, and dynamic stiffness. For the non-machined blades, i.e., blades before CNC machining was applied, static stiffness (Fig.~\ref{fig:NoMachined}(a)) is very similar for MS2D and the AUXETIC designs, around 7 N/µm. MS2R design and the solid one are the ones that present the lower stiffness value, near to zero, and, MS1D and MS1R are the ones with higher stiffness values. The dynamic stiffness of all blades is shown in Fig.~\ref{fig:NoMachined}(b). 
For the response, the reacceptance amplitude indicates that all blades are also very similar in terms of natural frequency (700 Hz) and damping (0.35$\%$).

For machined blades, the static stiffness, see Fig.~\ref{fig:Machined}(a), is very similar for MS1D, MS2D and the AUXETIC designs, around 6.5 N/µm. MS1R is the sample with higher stiffness value, around 10 N/µm, and, MS2R design and the solid blade are the ones with the lower values, near to zero. %Regarding dynamic stiffness, the reacceptance amplitude indicates that all blades are also very similar in terms of natural frequency (900 Hz) and damping (0.45$\%$).
In the context of dynamic stiffness analysis, receptance amplitude measurements revealed that the MS2R and MS2D designs exhibited lower natural frequencies (approximately 700 Hz) compared to the MS1D and MS1R designs (approximately 800 Hz), while maintaining similar damping ratios (0.45$\%$).

{\textit{Comparison against a solid blade.}} 
In relation to the solid blade, for both machined and no-machined workpieces, the static stiffness is around 0.2 N/µm and is the lowest of all. Regarding the dynamic behavior, the reacceptance amplitude of the response indicates that for no-machined solid samples, their results are similar to the MS2R and MS2D blade designs, while for machined specimens, there is a natural frequency of 966.25 Hz and 0.53$\%$ damping \cite{Marin-2024-SolidBlade}, which is higher than for microstructured blades.
When comparing the mass and the volume of the used material of the solid and the microstructured blades, the solid specimen weights 184 gr and occupy 23.5 mm$^3$, while the microstrucutred counterparts weight (in average) 136.7 gr and occupy 17.46 mm$^3$, which gives 26$\%$ weight reduction. The relative density of all microstructured blades is very similar, approximately $8.19 \times 10^{-6}$ Kg/mm$^3$. Observe that this result is not contrast to the simulation results in Section~\ref{ssec:OptNonHom} as there the weight reduction of the interior of $\Phi^{off}$ was estimated while here the weight reduction of the whole blade ($\Phi$) was measured.

%\mb{What is the difference machined-vs-nonmachined blades? Non-machined are those after LPBF?}

\begin{figure}[!tbh]
\hfill
  \begin{overpic}[width=0.88\columnwidth]{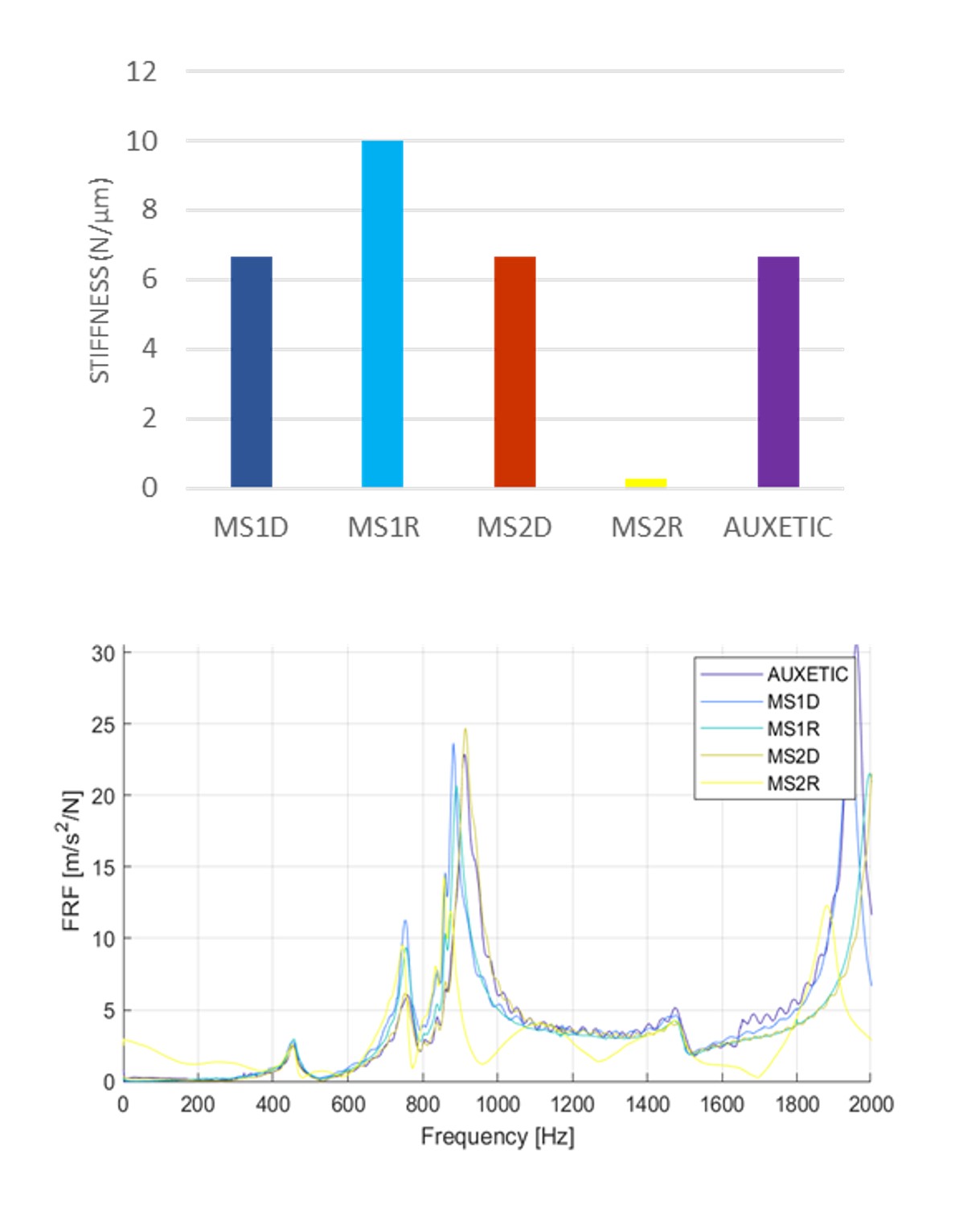}
  \put(0,92){(a)}
  \put(0,45){(b)}
  \end{overpic} 
\hfill \vrule width0pt\\
\vspace{-36pt}
	\caption{Hammer test for machined blades, after PBF-LB and machining manufacturing process (accelerometer positioned in 4 with hammer excitation in 2). (a) Measured static stiffness. (b) Measured receptance curves. }\label{fig:Machined}
\end{figure}

Finally, the experimentally measured frequencies exhibit good agreement with theoretical predictions.
The first natural frequency of the solid blade can be approximated by modeling the blade as a Bernoulli cantilever beam 
(recall~Fig.~\ref{fig:analysis:boundary_conditions}) with a length $L = 90\,\text{mm}$, a rectangular cross-section of width $b = 40\,\text{mm}$, and an average thickness $t = 6.75\,\text{mm}$.
Under this assumption, the first bending frequency is given by $f = 1.875^2 / 2\pi L^2 \sqrt{EI /\rho A}$, where $A = bt$ denotes the cross-sectional area, and $ I = bt^3/12 $ is the second moment of inertia.
By employing the material properties $E$ and $\rho$ defined in Section~\ref{sec:Analysis}, the computed natural frequency of the solid blade is approximately $677\,\text{Hz}$.
This value is slightly lower than the range of frequencies observed for the microstructured blades, which lie between $700$ and $800\,\text{Hz}$.

This discrepancy is consistent with theoretical expectations.
While the blade length $L$, Young’s modulus $E$, and density $\rho$ remain constant across different blade designs, variations in the cross-sectional area $A$ and moment of inertia $I$ affect the natural frequency through the factor $\sqrt{I/A}$.
Microstructured designs tend to yield a more substantial reduction in area (and hence mass) compared to inertia, resulting in an increase in $\sqrt{I/A}$, and therefore an increased frequency.

Specifically, designs such as MS1R and MS1D, which feature more severe material removal and thinner geometries, exhibit higher frequencies (approximately $800\,\text{Hz}$) compared to their thicker counterparts (approximately $700\,\text{Hz}$).
Assuming the reduction in inertia is negligible and considering an average area (and thus mass) reduction of approximately $26\%$, the expected frequency increase is about $16\%$, computed as $ \sqrt{1/(1 - 0.26)}$.
This yields a predicted frequency of approximately $785\,\text{Hz}$, which aligns well with the measured range for the microstructured blades.

{\textit{Tomography analysis}.}
To measure the quality of the microstructured blades, an X-Ray GE SEIFERT X-CUBE compact 195 KV tomography equipment was used, with a maximum resolution of 90 µm voxel size. This threshold includes tomography scan and image parameters in VGStudio MAX software. We measured the five specimen blades that contain different internal microstructures. The main objective was to measure the geometric deviations of internal microstructure produced during machining of the blades.

The deviations between the manufactured workpiece and the nominal model were measured. Each blade was fixed to the claw plate from its base and slightly inclined on the axis of rotation of the tomography, see Fig.~\ref{fig:InternalTomography} bottom-right. The scanning parameters were kept constant in all scans since all the pieces have very similar characteristics, namely, we set: 195 kV, 2.5 mA, 70 ms, 1 mm Cu filter and 0.5 mm Sn filter, using 130.9 mm voxel size.

\begin{figure*}[!tbh]
\hfill
  \begin{overpic}[width=0.98\textwidth]{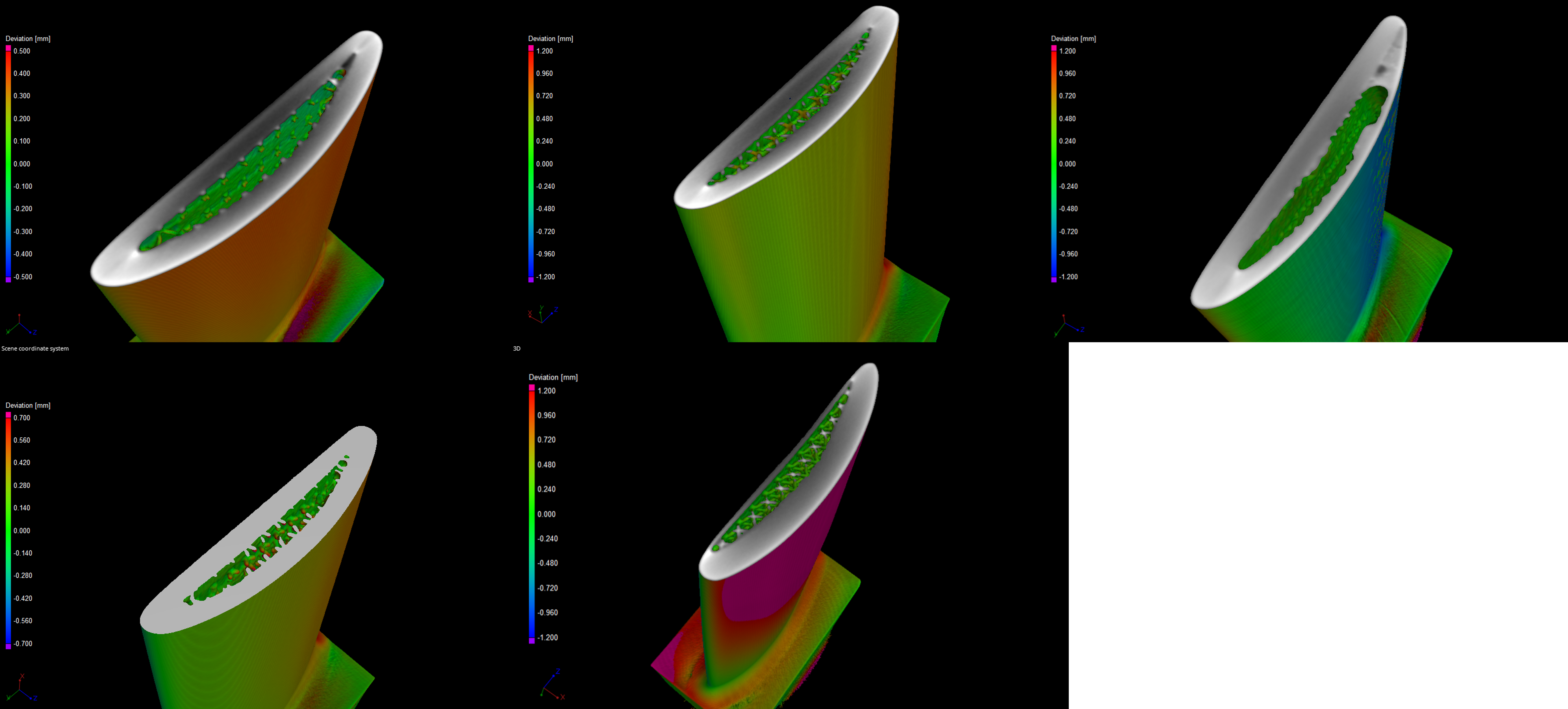}
  \put(75,1){\fcolorbox{gray}{white}{\includegraphics[angle=0,width=0.17\textwidth]{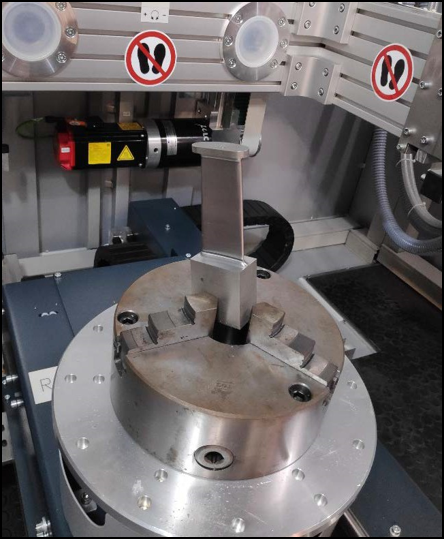}}}
  \put(5,42){\fcolorbox{gray}{white}{MS1D}}
  \put(5,20){\fcolorbox{gray}{white}{MS2D}}
  \put(40,42){\fcolorbox{gray}{white}{MS1R}}
  \put(40,20){\fcolorbox{gray}{white}{MS2R}}
  \put(73,42){\fcolorbox{gray}{white}{AUXETIC}}
  \end{overpic} 
\hfill \vrule width0pt\\
\vspace{-16pt}
	\caption{Tomography analysis of the internal part of the five sample designs (MS1D, MS1R, MS2D, MS2R and AUXETIC). Green color implies a very good alignment of CAD (STL) with printed structures. Bottom-right-framed: workpiece placement for tomography measurement.
    %\mb{The error bar is barely visible, had to scale it up, why do we have different viewpoints?}
    }\label{fig:InternalTomography}
\end{figure*}

The tomography measurements of the five specimen are shown in Figs.~\ref{fig:InternalTomography} and~\ref{fig:ExternalTomography}. 
The deviation analysis of the internal microstructure is shown in Fig.~\ref{fig:InternalTomography}, while the external  surface measurements are shown in Fig.~\ref{fig:ExternalTomography}. 
For all five test cases, the images show a deviation analysis and it can be seen that the deviation in the inner area is of a magnitude lower than the outer one. The maximum deviation in the inner area is: MS1D (±0.5 mm), MS1R (±1.2 mm), MS2D (±0.7 mm), MS2R (±1.2 mm), and AUXETIC (±1.2 mm). These values suggest that MS1R and MS2R were manufactured with good accuracy, while MS1D, MS2D, and AUXETIC have some internal inaccuracies. The green areas indicate a strong match between the CAD (STL) model and the printed structures.

It can be concluded from Fig.~\ref{fig:InternalTomography} that MS1R and MS2R present a well-manufactured internal microstructure, while the shape of internal designs MS1D, MS2D and AUXETIC were not correctly manufactured. The designed patterns present printing errors due to the bars dimensions and orientation, not completely compatible with PBF-LB geometry limitations.
%Auxetic one was the worst case, as shown in Fig.~\ref{fig:InternalTomography}. 
%\mb{Why is it worse, there is a lot of green???}

Fig.~\ref{fig:ExternalTomography} shows a comparison of the deviations on the surface of the externally visible face of the blades. MS1D and MS2D designs were the ones with smaller deviations (+0.4 mm). MS1R presents deviations around
+0.6 mm and MS2R around 0.8-1 mm. Finally, for the AUXETIC sample, the deviations were very high, as seen in the image.
%\mb{Again, I probably do not interpret correctly the plots, Auxetic does not seem to be the worst.}

% \begin{figure}[!tbh]
% \hfill
%   \begin{overpic}[width=0.65\columnwidth]{fig/5.png}
%   \end{overpic} 
% \hfill \vrule width0pt\\
% \vspace{-16pt}
% 	\caption{Workpiece placement for tomography measurement.}\label{fig:ComponentPosition}
% \end{figure}

%\begin{figure*}[!tbh]
%\hfill
%  \begin{overpic}[width=0.98\textwidth]{fig/6.png}
%  \put(77,1){\fcolorbox{gray}{white}{\includegraphics[angle=0,width=0.13\textwidth]{fig/5.png}}}
%  \end{overpic} 
%\hfill \vrule width0pt\\
%\vspace{-16pt}
%	\caption{Tomography analysis of the internal part of the five sample designs (MS1D, MS1R, MS2D, MS2R and AUXETIC). Green color implies a very good alignment of CAD (STL) with printed structures. Bottom-right-framed: workpiece placement for tomography measurement.
%    %\mb{The error bar is barely visible, had to scale it up, why do we have different viewpoints?}
%    }\label{fig:InternalTomography}
%\end{figure*}

\begin{figure}[!tbh]
\hfill
  \begin{overpic}[width=0.88\columnwidth]{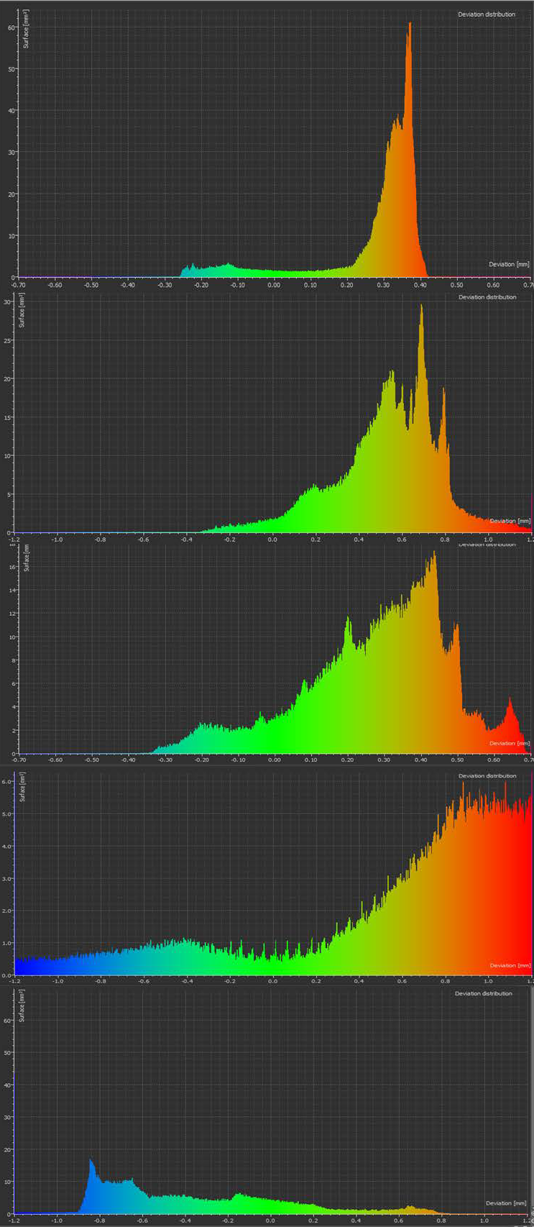}
  \put(5,95){\fcolorbox{gray}{white}{MS1D}}
  \put(5,72){\fcolorbox{gray}{white}{MS1R}}
  \put(5,51){\fcolorbox{gray}{white}{MS2D}}
  \put(5,33){\fcolorbox{gray}{white}{MS2R}}
  \put(5,15){\fcolorbox{gray}{white}{AUXETIC}}
  \end{overpic} 
\hfill \vrule width0pt\\
\vspace{-16pt}
	\caption{Tomography analysis of the external part of the sample designs (MS1D, MS1R, MS2D, MS2R and AUXETIC). %Left) Error measurement w.r.t to the nominal (CAD) geometry. Right) 
    The histograms show the deviation error (horizontal axis) and distribution of surface points with that error (vertical axis). Note that for formatting purposes, the histograms are scaled to the same width, however, the $x$ and $y$ scales are different.  The accuracy of the auxetic specimen was the least accurate, with many surface points with error close to 1 mm (blue).
    }\label{fig:ExternalTomography}
\end{figure}

%\pagebreak

%%%%%%%%%%%%%%%%%%%%%%%%%%%%%%%%%%%%%%%%%%%%%%%%%%%%%%%%%%%%%%%%%%%%%%%%%%%
\section{Discussion $\&$ Limitations}\label{sec:Limit}
%%%%%%%%%%%%%%%%%%%%%%%%%%%%%%%%%%%%%%%%%%%%%%%%%%%%%%%%%%%%%%%%%%%%%%%%%%%

\emph{Shape parameters.} Our pipeline contains several parameters, such as thickness or roundness, that control the shape of the interior tiles, recall Figs.~\ref{fig:GalleryOfTiles} and~\ref{fig-Diagonal-Tiles}. In the current implementation, these parameters are the same for all tiles in some examples, while in others (e.g., Fig.~\ref{fig-blade-lattice-cross}) the thicknesses of the tiles vary. Generalizing this setup such that each tile has its own set of shape parameters would be possible too (e.g.,~\cite{Zwar2022}), while the continuity across the tile boundaries should be preserved. Automating this optimization process with hundreds of thousands of degrees-of-freedom in one lattice is a worthy goal.  

\emph{Deformation of microstructures.} Based on the FEM results shown in Fig.~\ref{fig:analysis:results_graded_non_graded}, we can conclude that auxetic structures are not suited for this loading scenario and requirements, due to the large resulting deformations. However, it can be seen that spatial grading can reduce the deformations by a factor of two compared to a uniform layout (recall the deformation of the tip of the blade in Fig.~\ref{fig:analysis:results_graded_non_graded}).
It should be also noted that due to the large deformations, the assumptions of linear material and geometry behavior are invalid. For example, the rotational load is highly dependent on the state of deformation and therefore the loading will change as the geometry deforms. For small deformations, like in the case of the cross-like structure, this effect can be neglected. However, in the case of auxetic structures, the FEM results do not accurately reflect the real-world physical behavior and can therefore only be used for qualitative comparisons.

\emph{Extra material removal.} In our setup, the blades were placed vertically with holes at the top allowing to blow the extra material out. One could possibly look for different directions and/or placement to allow more effective extra material removal; we did not experiment with this setup.   

\emph{Heterogeneous materials.} Our test case specimen is a blisk blade that is typically manufactured using Inconel 718 alloy. We therefore considered only a single material. However, our pipeline supports the concept of heterogeneous materials and could consider various graded interior materials. 

\emph{Dynamical stiffness analysis.} 
Our results indicate that thinner structures demonstrate superior static and dynamic performance, when compared to microstructures with higher material volume or even the solid blade. This is attributed to their higher stiffness and lower susceptibility to machining-induced vibrations. The improved behavior is likely a result of the thinner structure’s ability to more effectively absorb and distribute cutting forces, thereby increasing overall stiffness and minimizing vibrational energy during the machining processes. These results are consistent with the hypothesis that reduced thickness contributes to higher structural stiffness and better stability during machining.

\emph{Topology optimization of bladed discs.} 
Topology optimization is an alternative approach to save material in the context of 3D printed light-weight structures. In the case of the aeronautical blisk, however, the state of the art are solid blades. There exist recent studies that consider topology-optimized bladed discs, see e.g.~\cite{Barreau-2022-TopOptBladedDisc}, which reports 
weight reduction between 15$\%$ and $32\%$. This data is roughly comparable to our results in terms of reduced mass/volume, 
yet \cite{Barreau-2022-TopOptBladedDisc} does not consider subsequent machining and optimized shapes as those shown in \cite[Fig.14]{Barreau-2022-TopOptBladedDisc} might be difficult to be CNC machined. Moreover, with holes throughout the whole blade, one would need to apply additional constraints to optimize only the interior of the blade. It might be an interesting venue for a future research though to further compare the proposed pipeline with topology optimization.

%%%%%%%%%%%%%%%%%%%%%%%%%%%%%%%%%%%%%%%%%%%%%%%%%%%%%%%%%%%%%%%%%%%%%%%%%%%properties
\section{Conclusion}\label{sec:Conclu}
%%%%%%%%%%%%%%%%%%%%%%%%%%%%%%%%%%%%%%%%%%%%%%%%%%%%%%%%%%%%%%%%%%%%%%%%%%%

We have introduced a new manufacturing paradigm for light-weight blade-like geometries, and demonstrated its feasibility by manufactured physical specimens. The proposed work considers the whole design-analysis-manufacturing-inspection process within a single framework and proposes microstructural design, that is optimized towards the functionality of the object under consideration (centrifugal force), and finally manufactured using hybrid manufacturing. We have shown that for the blisk blade geometry, one can significantly reduce the material without decreasing physical properties such as pressure sustainability.

The presented work serves as a proof of concept that one may design, optimize, manufacture and inspect microstructured objects that possess similar properties as their solid counterparts, while requiring considerably less material. As future research, one may further consider various free-form geometries, such as gears and/or georotors, or other engine components that are highly desirable to become lighter, e.g., in the new generation of electric cars. Another promising direction of future research is the use of multi-materials; currently, a homogeneous microstructure has been considered, however, it is expected, for example, that one could further reduce the object's weights when various{/graded} materials are considered. Finally, this academic work is paving the way towards possibly real industrial use of this effort, where the geometry as well as the topology of (individual) tiles could be optimized toward specific needs.

%%%%%%%%%%%%%%%%%%%%%%%%%%%%%%%%%%%%%%%%%%%%%%%%%%%%%%%%%%%%%%%%%%%%%%%%%%%%%%%%%%%%%%%%
\section*{Acknowledgments}
%%%%%%%%%%%%%%%%%%%%%%%%%%%%%%%%%%%%%%%%%%%%%%%%%%%%%%%%%%%%%%%%%%%%%%%%%%%%%%%%%%%%%%%%

This work was supported by the European Union's Horizon 2020 program under grant agreement No 862025.  

%Here we acknowledge our funding bodies. 

\bibliographystyle{elsarticle-num}
\bibliography{Machining,ADAM2,Auxetics}

\end{document}